\documentclass[aps,twocolumn,preprintnumbers,amsmath,amssymb,nofootinbib,superscriptaddress,notitlepage]{revtex4-1}

\usepackage{slashed}
\usepackage{soul}
\usepackage{mathrsfs,bm}
\usepackage{txfonts}
\usepackage{amsmath}
\usepackage{amssymb}
\usepackage{indentfirst}
\usepackage{graphicx,booktabs}
\usepackage{multirow}
\usepackage{overpic}
\usepackage{color}
\usepackage{amssymb}

\usepackage[utf8]{inputenc}

\makeatletter
\renewcommand{\p@subsection}{}
\renewcommand{\p@subsubsection}{}
\makeatother

\newcommand{\mpv}[1]{{#1}}

\begin{document}
\title{Five-flavor pentaquarks and other light- and heavy-flavor symmetry partners \\ of the LHCb hidden-charm pentaquarks}

\author{Fang-Zheng Peng}
\affiliation{School of Physics,  Beihang University, Beijing 100191, China}

\author{Ming-Zhu Liu}
\affiliation{School of Physics,  Beihang University, Beijing 100191, China}
\affiliation{School of Space and Environment, Beihang University, Beijing 100191, China}

\author{Ya-Wen Pan}
\affiliation{School of Physics,  Beihang University, Beijing 100191, China}

\author{Mario {S\'anchez S\'anchez}}
\affiliation{Centre d'\'Etudes Nucl\'eaires, CNRS/IN2P3, Universit\'e de Bordeaux, 33175 Gradignan, France}

\author{Manuel {Pavon} Valderrama}\email{mpavon@buaa.edu.cn}
\affiliation{School of Physics,  Beihang University, Beijing 100191, China}

\date{\today}
\begin{abstract}
  The discovery of three pentaquark peaks --- the $P_c(4312)$, $P_c(4440)$
  and $P_c(4457)$ --- by the LHCb collaboration
  has a series of interesting consequences
  for hadron spectroscopy.
  If {these hidden-charm objects} are indeed hadronic molecules, as suspected,
  they will be constrained by heavy-flavor and SU(3)-flavor
  symmetries.
  The combination of these two symmetries will imply the existence of
  a series of five-flavor pentaquarks with quark content
  $\bar{b} c s d u$ and $b \bar{c} s d u$, that is,
  pentaquarks that contain each of the five quark flavors that hadronize.
  In addition, from SU(3)-flavor symmetry alone we expect the existence of
  light-flavor partners of the three $P_c$ pentaquarks
  with strangeness $S=-1$ and $S=-2$.
  The resulting structure for the molecular pentaquarks is {analogous} 
  to the light{-}baryon octet --- we can label the pentaquarks
  as $P_{Q' \bar{Q}}^N$,
  $P_{Q' \bar{Q}}^{\Lambda}$, $P_{Q' \bar{Q}}^{\Sigma}$, $P_{Q' \bar{Q}}^{\Xi}$
  depending on their heavy- and light-quark content
  (with $N$, $\Lambda$, $\Sigma$, $\Xi$ the member of the light{-}baryon octet
  to which the light-quark structure resembles and $Q'$, $\bar Q$
  the heavy quark-antiquark pair).
  In total we predict $45$ new pentaquarks from heavy- and light-flavor
  symmetries alone, which extend up to $109$ undiscovered states
  if we also consider heavy-quark spin symmetry.
  If an isoquartet ($I={3/2}$) hidden-charm pentaquark is ever observed,
  this will in turn imply a second multiplet structure resembling
  the light{-}baryon decuplet: $P_{Q' \bar{Q}}^{\Delta}$,
  $P_{Q' \bar{Q}}^{\Sigma^*}$, $P_{Q' \bar{Q}}^{\Xi^*}$, $P_{Q' \bar{Q}}^{\Omega}$.
\end{abstract}


\maketitle

\section{Introduction}
The discovery by the LHCb collaboration of three
hidden{-}charm pentaquarks~\cite{Aaij:2019vzc} ---
the $P_c(4312)$, $P_c(4440)$ and $P_c(4457)$ ---
extends the previous observation of
the $P_c(4450)$ peak in 2015~\cite{Aaij:2015tga}.
Their masses and widths (in MeV) are
\begin{eqnarray}
  m_{P_{c1}}&=&4311.9\pm 0.7^{+6.8}_{-0.6} \, ,  \quad
  \Gamma_{P_{c1}}=9.8\pm2.7^{+3.7}_{-4.5} \, , \label{eq:m1} \\
  m_{P_{c2}}&=&4440.3\pm 1.3^{+4.1}_{-4.7} \, , \quad
  \Gamma_{P_{c2}}=20.6\pm4.9^{+8.7}_{-10.1} \, , \label{eq:m2} \\
  m_{P_{c3}}&=&4457.3\pm 0.6{}^{+4.1}_{-1.7} \, , \quad
  \Gamma_{P_{c3}}=6.4\pm2.0^{+5.7}_{-1.9} \, , \label{eq:m3}
\end{eqnarray}
where from now on we will use the notation $P_{c1}$, $P_{c2}$ and $P_{c3}$
for these three pentaquarks.
The $P_{c1}$ is $8.9\,{\rm MeV}$ below the $\bar{D} \Sigma_c$ threshold,
while the $P_{c2}$ and $P_{c3}$ are $21.8$ and $4.8\,{\rm MeV}$
below the $\bar{D}^* \Sigma_c$ threshold, respectively (where
we have considered these thresholds in the isospin{-}symmetric limit).
This, together with the existence of hidden-charm pentaquark predictions
in the molecular picture
before their experimental observation~\cite{Wu:2010jy,Wu:2010vk,Wu:2010rv,Xiao:2013yca,Karliner:2015ina,Wang:2011rga,Yang:2011wz},
{suggests} a molecular interpretation of these pentaquarks,
i.e. that they are bound states of a charmed antimeson
and a charmed baryon~\cite{Chen:2019bip,Chen:2019asm,Liu:2019tjn,Guo:2019fdo,Xiao:2019aya,Guo:2019kdc},
though this is not the only explanation that has been considered
by theoreticians~\cite{Eides:2019tgv,Wang:2019got,Cheng:2019obk}.

Heavy{-}hadron molecules are highly symmetrical: their light- and heavy-quark
content implies that they are constrained both by SU(3)-flavor
symmetry~\cite{GellMann:1962xb,Neeman:1961jhl}
and heavy-quark symmetry~\cite{Isgur:1989vq,Isgur:1989ed}.
Heavy-quark symmetry has in turn different manifestations, namely
heavy-quark spin symmetry (HQSS), heavy-flavor symmetry (HFS) 
and heavy{-}antiquark-diquark symmetry (HADS)~\cite{Savage:1990di},
which {altogether} provide deep insights
into the molecular spectrum~\cite{AlFiky:2005jd,Voloshin:2011qa,Mehen:2011yh,Valderrama:2012jv,Nieves:2012tt,HidalgoDuque:2012pq,Guo:2013sya,Guo:2013xga,Lu:2017dvm}.
The application of HQSS to the particular case of the LHCb pentaquarks
implies that the $P_{c1}$, $P_{c2}$ and $P_{c3}$ actually belong
to a multiplet composed of seven members~\cite{Liu:2019tjn,Xiao:2019aya,Sakai:2019qph,Yamaguchi:2019seo},
four of which {have} not been observed yet.
Before knowing that the $P_c(4450)$ peak contained two peaks,
HQSS was already used to predict a $J^P = 5/2^-$
$\bar{D}^* \Sigma_c^*$ molecular pentaquark
\mpv{and other partner states~\cite{Xiao:2013yca,Yamaguchi:2016ote,Yamaguchi:2017zmn,Shimizu:2018ran,Liu:2018zzu}.} 
In the past HFS and HADS have been applied to heavy meson-antimeson molecules
to explain spectroscopic relations among known molecular
states~\cite{Guo:2013sya} or to {deduce} the existence of
new states~\cite{Guo:2013xga}.
In this manuscript we will explore what are the consequences of
SU(3)-flavor symmetry and HFS if the hidden-charm pentaquarks
are indeed molecular.

\section{Symmetries}
First, we will consider the constraints that HFS and SU(3)-flavor symmetry
impose on the potential between a heavy antimeson and a heavy baryon.
HFS refers to the fact that the structure of a heavy-light
hadron (i.e. the ``brown muck'' around the heavy quark)
is independent of the flavor of the heavy quark.
As applied to heavy{-}hadron molecules, HFS implies that the potential
among heavy hadrons is independent of the flavor of
the heavy quarks inside the heavy hadrons.
The clearest example of this symmetry in molecular states are the
$Z_c$'s and $Z_b$'s resonances~\cite{Guo:2013sya},
which {are repeated} in the charm and bottom sectors and{ }are conjectured
to be $D^{(*)}\bar{D}^*$ and $B^{(*)}\bar{B}^*$ bound states, respectively.
If applied to the molecular pentaquarks, from HFS we expect
the potentials in the $\bar{D} \Sigma_c$, $\bar{D} \Sigma_b$,
${B} \Sigma_c$ and ${B} \Sigma_b$ two-body systems to be identical
(plus similar relations for the $\bar{D} \Sigma_c^*$,
$\bar{D}^* \Sigma_c$ and $\bar{D}^* \Sigma_c^*$ family of molecules).
For simplicity we will often use the generic notation $P$
and $P^*$ for the $J^P = 0^-, 1^-$ heavy mesons and
$\Sigma_Q$ and $\Sigma_Q^*$ for the $J^P = {1/2}^+$
and ${3/2}^+$ heavy baryons, irrespective of
whether they are their {charm} or bottom versions.
In addition we will use the notation $P_s$, $P_s^*$
for the heavy mesons with $S=1$ and $\Xi_Q'$, $\Xi_Q^*$
($\Omega_Q$, $\Omega_Q^*$) for the heavy baryons with $S=-1$ ($S=-2$).

\begin{table}[!ttt]
\begin{tabular}{|ccccc|}
\hline \hline
  Molecule & $I$ & $S$ & $V$ & $V_{\rm eigen}$ \\
  \hline
  $\bar{P} \Sigma_Q$ & $\tfrac{1}{2}$ & \phantom{+}$0$ & $V^O$ & $-$\\
  $\bar{P} \Sigma_Q$ & $\tfrac{3}{2}$ & \phantom{+}$0$ & $V^D$ & $-$\\
  $\bar{P} \Xi_Q'$ & $0$ & $-1$& $V^O$ & $-$\\
  $\bar{P} \Xi_Q'-\bar{P}_s \Sigma_Q$ & $1$ & $-1$ &
  $\begin{pmatrix}
    \frac{1}{3} V^O + \frac{2}{3} V^D & -\frac{\sqrt{2}}{3}\,(V^O - V^D) \\
    -\frac{\sqrt{2}}{3}\,(V^O - V^D) &  \frac{2}{3} V^O + \frac{1}{3} V^D \\ 
  \end{pmatrix}$ & $\{ V^O, V^D \}$ \\
  $\bar{P} \Omega_Q-\bar{P}_s \Xi_Q'$ & $\tfrac{1}{2}$ & $-2$ &
  $\begin{pmatrix}
    \frac{1}{3} V^O + \frac{2}{3} V^D & -\frac{\sqrt{2}}{3}\,(V^O - V^D) \\
    -\frac{\sqrt{2}}{3}\,(V^O - V^D) &  \frac{2}{3} V^O + \frac{1}{3} V^D \\ 
  \end{pmatrix}$ & $\{ V^O, V^D \}$ \\
  $\bar{P}_s \Omega_Q$ & $0$ & $-3$ & $V^D$ & $-$ \\
  \hline \hline 
\end{tabular}
\caption{The SU(3)-flavor structure of the potential for heavy meson-baryon
  molecules, where the heavy meson belongs to a SU(3)-flavor triplet and
  the heavy baryon to a sextet.
  The heavy meson-baryon potential can be decomposed into an octet and
  decuplet component, from which the octet piece corresponds to the
  potential for the hidden-charm molecular candidates.
  As a consequence other molecular pentaquarks belonging to the octet
  representation are also expected to bind.
  In addition to the SU(3)-flavor decomposition, the S-wave potential can
  be further decomposed into its light-quark structure,
  which is not explicitly shown here.
}
\label{tab:su3}
\end{table}

If we now consider SU(3)-flavor symmetry instead,
it happens that the $\bar{P}$, $\bar{P}_s$ heavy antimesons and
the $\Sigma_Q$, $\Xi_Q'$ and $\Omega_Q$ heavy baryons
belong to the $3$ and $6$ representation of
the SU(3)-flavor group, respectively~\footnote{
We will not consider explicitly the difference between ground{-}
and excited{-}state heavy hadrons, as it does not affect their
light-flavor structure.}.
Two-body heavy antimeson-baryon states can be decomposed
into $3 \otimes 6 = 8 \oplus 10$, i.e. into the octet
and decuplet representations, where the SU(3)
Clebsch-Gordan coefficients can be consulted
in Ref.~\cite{Kaeding:1995vq}.
\mpv{This octet and decuplet decomposition is not dependent on the nature of
  the pentaquarks, but on their light-quark content, and it has indeed been
  previously pointed out for compact pentaquarks~\cite{Santopinto:2016pkp}.
  Within the molecular explanation, this decomposition specifically
}
implies that the heavy antimeson-baryon potential
can be decomposed into a linear combination of an octet
and decuplet contribution
\begin{eqnarray}
  V = \lambda^O V^O + \lambda^D V^D \, ,
\end{eqnarray}
with $V^O$ and $V^D$ the octet and decuplet pieces
and $\lambda^O$, $\lambda^D$ numerical coefficients.
We show the full decomposition in Table \ref{tab:su3},
which happens to be surprisingly simple:
for most heavy antimeson-baryon molecules,
the potential is a pure octet or decuplet contribution.
In turn, this is easily explained from the observation that the resulting
pentaquarks have the same quantum numbers as the corresponding octet or
decuplet light baryons.
Even for the $\bar{P} \Xi_Q'-\bar{P}_s \Sigma_Q$ and
$\bar{P} \Omega_Q-\bar{P}_s \Xi_Q'$ molecules (where the dash indicates
that these channels couple), for which the potential is a $2 \times 2$ matrix,
when we look at the eigenvalues we recover
\begin{eqnarray}
  V = \lbrace V^O, V^D\rbrace \, ,
\end{eqnarray}
depending on the linear combination of the two channels, with the octet
eigenvalue corresponding to
\begin{eqnarray}
  | 8 \rangle &=& -\sqrt{\frac{1}{3}}\,| \bar{P} \Xi_Q'(I=1) \rangle +
  \sqrt{\frac{2}{3}}\,| \bar{P}_s \Sigma_Q \rangle \, , \\
  | 8 \rangle &=& -\sqrt{\frac{1}{3}}\,| \bar{P} \Omega_Q \rangle +
  \sqrt{\frac{2}{3}}\,| \bar{P}_s \Xi_Q' \rangle \, ,
\end{eqnarray}
and the decuplet eigenvalue to
\begin{eqnarray}
  | 10 \rangle &=& \sqrt{\frac{2}{3}}\,|\bar{P} \Xi_Q'(I=1) \rangle +
  \sqrt{\frac{1}{3}}\,|\bar{P}_s \Sigma_Q \rangle \, , \\
  | 10 \rangle &=& \sqrt{\frac{2}{3}}\,|\bar{P} \Omega_Q \rangle +
  \sqrt{\frac{1}{3}}\,|\bar{P}_s \Xi_Q' \rangle \, .
\end{eqnarray}
These two molecular systems, $\bar{P} \Xi_Q'-\bar{P}_s \Sigma_Q$ and
$\bar{P} \Omega_Q-\bar{P}_s \Xi_Q'$, will adopt the lowest{-}energy
configuration, be it either the octet or decuplet one.
In the absence of additional experimental information and 
knowing that the $P_{c1}$, $P_{c2}$ and $P_{c3}$ hidden-charm pentaquarks
most probably belong to the octet, we naively expect
the lowest{-}energy configuration to be the octet~\footnote{We notice
  that a recent work~\cite{Meng:2019fan} has predicted a series of
  $c\bar{c} sss$ ($P_c^{\Omega}$) pentaquarks (but compact, instead of molecular).
  This suggests that a few of the decuplet configurations
  might be attractive as well.}.

Owing to heavy-flavor symmetry, the potential is expected to be
independent of the flavor of the heavy{ }quarks. This implies
in particular that the octet configurations
\begin{eqnarray}
  \bar{D} \Xi_b'(I=0)\,, & \quad
  \bar{D} \Xi_b'(I=1)-\bar{D}_s \Sigma_b \, , \\
  {B} \Xi_c'(I=0)\,, & \quad
  {B} \Xi_c'(I=1)-{B}_s \Sigma_c \, ,
\end{eqnarray}
which contain the five quark flavors that hadronize,
will display as much attraction as the hidden-charm pentaquarks.
Out of the four five-flavor configurations, the strange-isoscalar
molecules [$\bar{D} \Xi_b'(0)$, $B \Xi_c'(0)$]
are relatively easy to deal with (they are single{-}channel systems).
For the strange-isovector molecules [$\bar{D} \Xi_b'(I=1)-\bar{D}_s \Sigma_b$, ${B} \Xi_c'(I=1)-{B}_s \Sigma_c$] we have a two-channel problem
where the thresholds are separated by about $20\,{\rm MeV}$
and $40\,{\rm MeV}$ for the isovector ${\bar b} c s q q$ and
${b} {\bar c} s q q$ pentaquark configurations, respectively.
The question is whether this energy gap will prevent a predominantly octet
molecular state to form or not.
The answer depends on the comparison of the momentum scales of
the binding mechanism and the coupled-channel dynamics.
The typical momentum scale of the coupled channels~\footnote{
  This momentum scale is defined as $\sqrt{2 \mu \Delta}$,
  with $\mu$ the reduced mass of the system and
  $\Delta$ the mass gap between the channels.}
in the previous cases
is about $250\,{\rm MeV}$ and $350\,{\rm MeV}$ for the ${\bar b} c s u d$
and ${b} {\bar c} s u d$ pentaquarks, while the binding mechanism is
expected to be short-ranged (e.g. vector{-}meson exchange),
with a momentum scale of the order of
$(0.5-1.0)\,{\rm GeV}$ give or take.
As a consequence, we expect the isovector five-flavor pentaquarks to bind
(a conjecture which we confirm by means of concrete calculations
in what follows).

\section{Effective field theory description}
\mpv{
  To explicitly check the effects of the previous symmetries, we will describe
  the pentaquarks as non-relativistic meson-baryon bound states interacting
  by means of a contact-range potential that is heavy- and
  SU(3)-flavor symmetric.

  This choice is not arbitrary, but corresponds with the lowest or leading
  order (${\rm LO}$) effective field theory (EFT) description of the heavy
  antimeson and heavy baryon two-body system.
  EFTs exploit the existence of a separation of scales to formulate generic
  low energy descriptions of physical systems. The idea is to identify
  characteristic low and high energy scales $Q$ and $M$ such that
  $Q/M \ll 1$ and then express every physical quantity
  as a power series in terms of the ratio $Q/M$.
  The first term in this series is the LO, the second is
  the next-to-leading order (${\rm NLO}$), and so on.
  
  For molecular pentaquarks the required scale separation manifest itself as
  follows:
  the typical low energy scale $Q$ is of the order of $(100-200)\,{\rm MeV}$ and
  can be identified with the pion mass or the binding momentum of
  the pentaquarks.
  At this scale the meson-baryon dynamics is well known and involves
  the exchanges of pions and other pseudoscalar mesons.
  The high energy scale $M$ is in the $(0.5-1.0)\,{\rm GeV}$ range and
  can be identified with the rho meson mass or the momentum scale
  at which the internal structure of the hadrons becomes evident.
  This part of the interaction is less well-known and might very well involve
  non-molecular components of the pentaquark wave function.
  EFT parametrizes it as a series of contact-range operators.
  
  Our ${\rm LO}$ description of the pentaquarks only involves
  the contact-range potential.
  This choice is justified (i) from a well-known EFT observation
  that indicates that the existence of shallow bound states
  (e.g. the deuteron or near-threshold states such as
  hadronic molecules) increases the importance of contact-range interactions
  at low energies~\cite{vanKolck:1998bw,Chen:1999tn} and
  (ii) from concrete EFT calculations
  for the LHCb pentaquarks that suggest that pion
  exchanges are ${\rm NLO}$ and thus a perturbative
  correction to the LO results~\cite{PavonValderrama:2019nbk}.
}

\mpv{From the previous, the ${\rm LO}$} S-wave interaction binding
the $P_{c1}$, $P_{c2}$ and $P_{c3}$ molecular pentaquarks will be
{given by the Lagrangian
  \begin{eqnarray}
    \mathcal{L}_{\rm contact} = C_i^O \sum_{IS}
    (o_{IS}^{abc} M_a P_i^{J} B_{bc})^{\dagger}
    (o_{IS}^{abc} M_a P_{i}^{J} B_{bc}) \, ,
    \label{eq:L-contact}
  \end{eqnarray}
  where $C^O_i$ is the (octet) coupling constant, $i=1,2,3$ is the index
  with which we label the hidden-charm pentaquarks,
  $M_a$ is a triplet heavy meson
  with the quark content $|\bar{Q} q_a \rangle$, where $q_a = u, d, s$
  depending on the flavor index $a$, $B_{bc}$ a sextet heavy baryon with quark
  content $|{Q} \frac{1}{\sqrt{2}} (q_b q_c + q_c q_b) \rangle$
  (i.e. symmetric in the flavor indices), $o_{IS}^{abc}$ is a tensor
  in flavor space that projects the heavy antimeson-baryon system
  in an octet state with given isospin $I$ and strangeness $S$
  (the exact form of this tensor can be deduced from Table ~\ref{tab:su3}), 
  and $P_i^{J}$ is a projector into the corresponding
  spin channel $J$ if there is more than
  one~\footnote{The form of this projector is trivial ($P_1 = 1$) for the
    $P_{c1}$ pentaquark, while $P_2$ and $P_3$ depend on the spin of
    the $P_{c2}$ and $P_{c3}$ pentaquarks, which is either
    $J = \tfrac{1}{2}$ or $\tfrac{3}{2}$, where the projector
    for the $| J M \rangle$ spin configuration
    in the $\bar{D}^* \Sigma_c$ system takes the form
    $\langle 1 m_1 | P_{J M} | \tfrac{1}{2} m_2 \rangle =
    \langle 1 m_1 \tfrac{1}{2} m_2 | J M \rangle$, i.e. it coincides
    with the Clebsch-Gordan coefficients coupling a $\bar{D}^*$ meson
    and $\Sigma_c$ baryon with spin wave functions $| 1 m_1 \rangle$
    and $|\tfrac{1}{2} m_2 \rangle$ to total spin $| J M \rangle$.
  }.
  For molecular pentaquarks, the spin of the $P_{c1}$
  will be $J = \tfrac{1}{2}$, while for the $P_{c2}$ and $P_{c3}$
  it will be either $J = \tfrac{1}{2}$ or $\tfrac{3}{2}$,
  though we do not know which of these two pentaquarks corresponds to
  each of the two possible spin configurations.
  We are also assuming that the decuplet contact-range interaction is
  subleading, which is why it is not included in the Lagrangian above.
}

The previous Lagrangian generates a simple contact-range potential of the type
\begin{eqnarray}
  \langle p' | V | p \rangle = C^O_i(\Lambda)\,
  f(\frac{p}{\Lambda})\,f(\frac{p'}{\Lambda}) \, ,
  \label{eq:contact}
\end{eqnarray}
where we have regularized the potential, originally a Dirac{ }delta
in momentum space, with the Gaussian regulator  $f(x) = e^{-x^2}$
and a cutoff $\Lambda$.
For the cutoff we choose the range $\Lambda = (0.5-1.0)\,{\rm GeV}$,
i.e. around the $\rho$ meson mass.
With this potential we solve a coupled-channel Lippmann-Schwinger
equation of the type
\begin{eqnarray}
  \phi_A(k) + \sum_B \int \frac{d^3 p}{(2\pi)^3}\,\langle k | V_{AB} | p \rangle
  \,\frac{\phi_B(p)}{M_B + p^2/(2 \mu_B) - M_P} = 0,\notag\\ \, 
  \label{eq:BSE}
\end{eqnarray}
where $A,B$ are indices for the channels we are considering,
$\phi_A$ the vertex function
(i.e. the wave function $\Psi_A$ times the propagator,
  $\phi_A(p) = [M_A + p^2/(2\mu_A) - M_P]\,\Psi_A(p)$),
$V_{AB}$ the potential between
channels $A$ and $B$, $M_B$ the total mass of
the heavy antimeson and baryon comprising channel $B$,
$\mu_B$ their reduced mass and $M_P$ the mass of
the molecular pentaquark we are predicting.
We notice that the only configurations with more than one channel
are the $(I,S) = (1,-1)$ and $(\tfrac{1}{2},-2)$, see Table \ref{tab:su3}.
For illustrative purposes we consider the bound-state equation
for a Gaussian regulator in the single-channel case,
in which it reduces to
\begin{equation}
  1 + C_i^O(\Lambda)\,\frac{\mu_A}{4 \pi^2}\,I_0(\gamma_A, \Lambda) = 0 \, ,
\end{equation}
with $\gamma_A = \sqrt{2\mu_A(M_A-M_P)}$ the wave number of
the molecular pentaquark and where $I_0$ is given by
\begin{equation}
  I_0(\gamma_A, \Lambda) = \sqrt{2 \pi}\,\Lambda -
  2\, e^{2 \gamma_A^2 / \Lambda^2}\,\pi \gamma_A \,
  {\rm erfc}\left( \frac{\sqrt{2} \gamma_A}{\Lambda} \right) \, ,
\end{equation}
where ${\rm erfc}\,(x)$ is the complementary error function.

If we determine the $C_i^O$ couplings from reproducing the masses of
the $i = 1,2,3$ $P_{ci}$ pentaquark, for $\Lambda = 0.75\,{\rm GeV}$
we obtain the couplings
\begin{eqnarray}
  C^O_1 &=& -1.19\,(-(2.17-0.80))\,{\rm fm}^2 \, , \\
  C^O_2 &=& -1.44\,(-(2.88-0.93))\,{\rm fm}^2 \, , \\
  C^O_3 &=& -1.02\,(-(1.80-0.71))\,{\rm fm}^2 \, ,  
\end{eqnarray}
where the values in parentheses correspond to varying the cutoff
in the $(0.5-1.0)\,{\rm GeV}$ window~\footnote{For simplicity,
  we have not considered the errors esteeming from the uncertainties
  in the pentaquark masses, see Eqs.~(\ref{eq:m1}-\ref{eq:m3}),
  nor from the further dependence of these masses
  on the resonance profile, check for instance
  Ref.~\cite{Fernandez-Ramirez:2019koa} in which
  the $P_c(4312)$ is found to be a virtual (instead of a bound) state.}.
With these couplings, for $\Lambda = 0.75\,{\rm GeV}$ we predict
the location of the ${\bar c} {b}$ five-flavor pentaquarks to be
\begin{eqnarray}
  m(P^{\Lambda}_{\bar{c} b}) &=&
  7783^{+6}_{-5} \, , \, 7907 \pm 7 \, , \, 7930^{+5}_{-4}  \,{\rm MeV} \, ,
  \label{eq:P5-a} \\
  m(P^{\Sigma}_{\bar{c} b}) &=&
  7765^{+6}_{-5} \, , \, 7892^{+8}_{-9} \, , \,  7914^{+5}_{-4}  \,{\rm MeV} \, , 
\end{eqnarray}
where the uncertainty comes from varying the cutoff
(i.e. taking $\Lambda = (0.5-1.0)\,{\rm GeV}$), {but does not include
the SU(3) symmetry breaking effects, which we discuss later}.
For the $c {\bar b}$ five-flavor pentaquarks we predict instead
\begin{eqnarray}
  m(P^{\Lambda}_{{c} {\bar b}}) &=&
  7829^{+10}_{-9} \, , \, 7858^{+12}_{-10}\, , \,  7883^{+8}_{-7}  \,{\rm MeV} \, , \\
  m(P^{\Sigma}_{{c} {\bar b}}) &=&
  7804^{+6}_{-5} \, , \, 7835^{+8}_{-7} \, , \,  7858^{+5}_{-4}  \,{\rm MeV} \, .
  \label{eq:P5-d}
\end{eqnarray}
The complete list of predictions (including not only cutoff 
but also SU(3)-flavor uncertainties) can be consulted
in Table \ref{tab:partners}.

\begin{table*}[!ttt]
\begin{tabular}{|cccccc|cccccc|}
\hline \hline
Molecule & $I$ & $S$ & $B_P$ & $M_P$ & Partner
&
Molecule & $I$ & $S$ & $B_P$ & $M_P$ & Partner 
\\
  \hline
  $\bar{D} \Sigma_c$ & $\tfrac{1}{2}$ & \phantom{+}$0$ & Input & Input
  & $P_{c1}$
  &
  ${B} \Sigma_c$ & $\tfrac{1}{2}$ & \phantom{+}$0$ &
  $27.5^{+9.5}_{-8.0}$ & $7710.5^{+8.0}_{-9.5}$ & $P_{c1}$
  \\
  $\bar{D}^* \Sigma_c$ & $\tfrac{1}{2}$ & \phantom{+}$0$ & Input & Input
  & $P_{c2}$
  &
  ${B}^* \Sigma_c$ & $\tfrac{1}{2}$ & \phantom{+}$0$ &
  $43.6^{+10.6}_{-9.3}$ & $7734.6^{+9.3}_{-10.6}$
  & $P_{c2}$
  \\
  $\bar{D}^* \Sigma_c$ & $\tfrac{1}{2}$ & \phantom{+}$0$ & Input & Input
  & $P_{c3}$
  &
  ${B}^* \Sigma_c$ & $\tfrac{1}{2}$ & \phantom{+}$0$ &
  $18.6^{+7.6}_{-6.0}$ & $7759.7^{+6.0}_{-7.6}$
  & $P_{c3}$
  \\
  \hline
  $\bar{D} \Xi_c'$ & $0$ & $-1$ &
  $9.6^{+10.4}_{-7.3}$ & $4436.3^{+7.3}_{-10.4}$ & $P_{c1}$
  &
  ${B} \Xi_c'$ & $0$ & $-1$ &
  $29^{+18}_{-16}$ & $7829^{+16}_{-18}$ & $P_{c1}$
  \\
  $\bar{D}^* \Xi_c'$ & $0$ & $-1$ &
  $23^{+16}_{-13}$ & $4565^{+13}_{-16}$ & $P_{c2}$
  &
  ${B}^* \Xi_c'$ & $0$ & $-1$ &
  $45^{+23}_{-21}$ & $7858^{+21}_{-23}$ & $P_{c2}$ 
  \\
  $\bar{D}^* \Xi_c'$ & $0$ & $-1$ &
  $5.4^{+7.7}_{-4.7}$ & $4581.8^{+4.7}_{-7.7}$ & $P_{c3}$
  &
  ${B}^* \Xi_c'$ & $0$ & $-1$ &
  $20^{+15}_{-12}$ & $7883^{+12}_{-15}$ & $P_{c3}$
  \\
  \hline
  $\bar{D} \Xi_c'-\bar{D}_s \Sigma_c$ & $1$ & $-1$ &
  $5.2^{+9.4}_{-5.0}$ & $4416.7^{+5.0}_{-9.4}$ &
  $P_{c1}$ &
  ${B} \Xi_c'-{B}_s \Sigma_c$  & $1$ & $-1$ &
  $20^{+17}_{-14}$ & $7801^{+14}_{-17}$ & $P_{c1}$
  \\
  $\bar{D}^* \Xi_c'-\bar{D}_s^* \Sigma_c$ & $1$ & $-1$ &
  $18^{+16}_{-12}$ & $4548^{+12}_{-16}$ & $P_{c2}$
  &
  ${B}^* \Xi_c'-{B}_s^* \Sigma_c$ & $1$ & $-1$ &
  $36^{+22}_{-19}$ & $7833^{+19}_{-22}$ & $P_{c2}$
  \\
  $\bar{D}^* \Xi_c'-\bar{D}_s^* \Sigma_c$ & $1$ & $-1$ &
  $2.0^{+6.5}_{-2.0}$ & $4563.7^{+2.0}_{-6.5}$
  & $P_{c3}$ &
  ${B}^* \Xi_c'-{B}_s^* \Sigma_c$ & $1$ & $-1$ &
  $12^{+13}_{-10}$ & $7857^{+10}_{-13}$ & $P_{c3}$
  \\
  \hline
  $\bar{D} \Omega_c-\bar{D}_s \Xi_c'$ & $\tfrac{1}{2}$ & $-2$ &
  $2.6^{+9.4}_{-2.6}$ & $4544.2_{-9.4}^{+2.6}$ & $P_{c1}$
  &
  ${B} \Omega_c-{B}_s \Xi_c'$ & $\tfrac{1}{2}$ & $-2$
  & $14^{+17}_{-13}$ & $7931^{+13}_{-17}$ & $P_{c1}$
  \\
  $\bar{D}^* \Omega_c-\bar{D}_s^* \Xi_c'$  & $\tfrac{1}{2}$ & $-2$
  & $16^{+16}_{-13}$ & $4675^{+13}_{-15}$ & $P_{c2}$
  &
  ${B}^* \Omega_c-{B}_s^* \Xi_c'$  & $\tfrac{1}{2}$ & $-2$
  & $31^{+23}_{-20}$ & $7963^{+23}_{-20}$ & $P_{c2}$
  \\
  $\bar{D}^* \Omega_c-\bar{D}_s^* \Xi_c'$ & $\tfrac{1}{2}$ & $-2$
  & $0.4^{+6.2}_{-0.4}$ & $4690.3_{-6.2}^{+0.4}$ & $P_{c3}$
  &
  ${B}^* \Omega_c-{B}_s^* \Xi_c'$ & $\tfrac{1}{2}$ & $-2$
  & $7.3^{+13.0}_{-8.2}$ & $7986.5^{+8.2}_{-13.0}$ & $P_{c3}$
  \\
  \hline 
  \hline
  $\bar{D} \Sigma_b$ & $\tfrac{1}{2}$ & \phantom{+}$0$
  & $20.2^{+5.3}_{-4.7}$ & $7660.1^{+4.7}_{-5.3}$ & $P_{c1}$
  &
  ${B} \Sigma_b$ & $\tfrac{1}{2}$ & \phantom{+}$0$
  & $48^{+23}_{-18}$ & $11044^{+18}_{-23}$ & $P_{c1}$
  \\
  $\bar{D}^* \Sigma_b$ & $\tfrac{1}{2}$ & \phantom{+}$0$
  & $37.5^{+7.3}_{-6.5}$ & $7784.2^{+6.5}_{-7.3}$ & $P_{c2}$
  &
  ${B}^* \Sigma_b$ & $\tfrac{1}{2}$ & \phantom{+}$0$
  & $68^{+25}_{-28}$ & $11070^{+28}_{-25}$ & $P_{c2}$
  \\
  $\bar{D}^* \Sigma_b$ & $\tfrac{1}{2}$ & \phantom{+}$0$
  & $14.3^{+4.7}_{-4.0}$ & $7807.4^{+4.0}_{-4.7}$ & $P_{c3}$
  &
  ${B}^* \Sigma_b$ & $\tfrac{1}{2}$ & \phantom{+}$0$
  & $37^{+19}_{-15}$ & $11101^{+15}_{-19}$ & $P_{c3}$
  \\
  \hline
  $\bar{D} \Xi_b'$ & $0$ & $-1$
  & $20^{+15}_{-12}$ & $7783^{+12}_{-15}$ & $P_{c1}$
  &
  ${B} \Xi_b'$ & $0$ & $-1$
  & $49^{+29}_{-25}$ & $11166^{+25}_{-29}$ & $P_{c1}$
  \\
  $\bar{D}^* \Xi_b'$ & $0$ & $-1$
  & $38^{+20}_{-18}$ & $7907^{+18}_{-20}$ & $P_{c2}$
  &
  ${B}^* \Xi_b'$ & $0$ & $-1$
  & $68^{+34}_{-30}$ & $11192^{+34}_{-30}$ & $P_{c2}$
  \\
  $\bar{D}^* \Xi_b'$ & $0$ & $-1$
  & $15^{+12}_{-10}$ & $7930^{+10}_{-12}$ & $P_{c3}$
  &
  ${B}^* \Xi_b'$ & $0$ & $-1$
  & $38^{+25}_{-21}$ & $11222^{+21}_{-25}$ & $P_{c3}$
  \\
  \hline
  $\bar{D} \Xi_b'-\bar{D}_s \Sigma_b$ & $1$ & $-1$
  & $16^{+14}_{-12}$ & $7765^{+12}_{-14}$ & $P_{c1}$
  &
  ${B} \Xi_b'-{B}_s \Sigma_b$  & $1$ & $-1$
  & $40^{+28}_{-24}$ & $11140^{+24}_{-28}$ & $P_{c1}$
  \\
  $\bar{D}^* \Xi_b'-\bar{D}_s^* \Sigma_b$ & $1$ & $-1$
  & $34^{+20}_{-18}$ & $7892^{+18}_{-20}$ & $P_{c2}$
  &
  ${B}^* \Xi_b'-{B}_s^* \Sigma_b$  & $1$ & $-1$
  & $59^{+33}_{-29}$ & $11161^{+29}_{-33}$ & $P_{c2}$
  \\
  $\bar{D}^* \Xi_b'-\bar{D}_s^* \Sigma_b$ & $1$ & $-1$
  & $11^{+12}_{-10}$ & $7914^{+12}_{-10}$ & $P_{c3}$
  &
  ${B}^* \Xi_b'-{B}_s^* \Sigma_b$  & $1$ & $-1$
  & $30^{+26}_{-19}$ & $11199^{+26}_{-19}$ & $P_{c3}$
  \\
  \hline
  $\bar{D} \Omega_b-\bar{D}_s \Xi_b'$ & $\tfrac{1}{2}$ & $-2$
  & $15^{+15}_{-12}$ & $7888^{+12}_{-15}$ & $P_{c1}$
  &
  ${B} \Omega_b-{B}_s \Xi_b'$  & $\tfrac{1}{2}$ & $-2$
  & $35^{+29}_{-24}$ & $11267^{+24}_{-29}$ & $P_{c1}$
  \\
  $\bar{D}^* \Omega_b-\bar{D}_s^* \Xi_b'$  & $\tfrac{1}{2}$ & $-2$
  & $34^{+20}_{-18}$ & $8013^{+18}_{-20}$ & $P_{c2}$
  &
  ${B}^* \Omega_b-{B}_s^* \Xi_b'$  & $\tfrac{1}{2}$ & $-2$
  & $56^{+34}_{-29}$ & $11295^{+30}_{-34}$ & $P_{c2}$
  \\
  $\bar{D}^* \Omega_b-\bar{D}_s^* \Xi_b'$ & $\tfrac{1}{2}$ & $-2$
  & $11^{+12}_{-9}$ & $8037^{+9}_{-12}$ & $P_{c3}$
  &
  ${B}^* \Omega_b-{B}_s^* \Xi_b'$  & $\tfrac{1}{2}$ & $-2$
  & $26^{+24}_{-19}$ & $11325^{+19}_{-24}$ & $P_{c3}$
  \\
  \hline \hline
\end{tabular}
\caption{
  The heavy- and light-flavor symmetry partners of
  the LHCb pentaquark trio, the $P_c(4312)$, $P_c(4440)$ and
  $P_c(4457)$ (or $P_{c1}$, $P_{c2}$, $P_{c3}$ for short).
  This includes the five-flavor pentaquarks with quark content
  ${\bar b} c s d u$ and $b {\bar c} s d u$.
  The column ``Molecule'' displays the two-hadron system under consideration,
  $I$ the isospin, $S$ the strangeness, $B_P$ the binding energy,
  $M_P$ the mass (where $M_P = M_{\rm th} - B_P$, with $M_{\rm th}$ the mass of
  the corresponding heavy antimeson-baryon threshold, for which we take
  the isospin symmetric limit of the masses listed in the Review of Particle
  Physics (RPP)~\cite{Zyla:2020zbs})
  and ``Partner'' represents which hidden-charm
  pentaquark ($P_{ci}$, $i=1,2,3$) is the partner of the predicted state.
  In the coupled-channel cases, the binding energy is calculated
  relative to the channel with the lowest mass.
  For the calculations we use a contact-range EFT, with the potential
  of Eq.~(\ref{eq:contact}) and a Gaussian regulator
  with a cutoff $\Lambda = 0.75\,{\rm GeV}$.
  The error comes from two different sources, which are added
  in quadrature: (i) varying the cutoff in the $\Lambda = (0.5-1.0)\,{\rm GeV}$
  range and (ii) assuming a $20\%$ uncertainty in SU(3)-flavor symmetry
  as applied to the contact-range couplings (this second error
  only pertains pentaquarks with strangeness).
  In general the SU(3)-flavor uncertainty dominates in the $c\bar{c}$,
  $c \bar{b}$, $b \bar{c}$ sectors, while for the $b\bar{b}$ pentaquarks
  the bulk of the errors come from the cutoff variation (in agreement
  with theoretical expectations~\cite{Baru:2018qkb}).
}
\label{tab:partners}
\end{table*}

The spectrum of Table \ref{tab:partners} implies that each of the observed
hidden-charm pentaquarks belongs to a light/heavy-flavor multiplet
with $16$ members.
As three hidden-charm pentaquarks have been observed,
this means a total of $48$ states (of which $45$ are so far unobserved).
The experimental observation of these pentaquarks could be achieved by means of
the SU(3)-flavor and HFS analogues of the $J/\Psi N$ decay channel
that has been used in the discovery of the $P_{c1}$, $P_{c2}$
and $P_{c3}$.
For instance, the five-flavor pentaquarks $P^{\Lambda}_{c \bar b}$
and $P^{\Sigma}_{c \bar b}$ could be detected by means of
their $B_c^{+} \Lambda$ and $B_c^{+} \Sigma$ decays.

Even though for the moment we have not considered HQSS explicitly ,
it is easy to figure out its consequences: from HQSS we expect
\mpv{the hidden-charm pentaquarks to come in multiplets of
  up to seven members~\cite{Xiao:2013yca,Yamaguchi:2016ote,Yamaguchi:2017zmn,Shimizu:2018ran}.
  Within the scope of contact-range EFTs incorporating HQSS~\cite{Liu:2018zzu},
  the observation of the $P_{c1}$, $P_{c2}$ and $P_{c3}$ pentaquarks suggests
  that the aforementioned septuplet is probably
  complete~\cite{Liu:2019tjn,Du:2019pij},
  meaning that there are $4$ unobserved states.
  This result is reproduced in most schemes that include HQSS,
  e.g. models with a compact core coupled to the molecular
  degrees of freedom~\cite{Yamaguchi:2019seo}, indicating
  that it depends on HQSS instead of the specific
  dynamics generating the pentaquarks.
  The bottom-line is that
}
if we compound the HQSS multiplets with the SU(3)-flavor and HFS ones,
the heavy molecular pentaquark family could contain a total of
$112$ states ($3$ observed, $109$ to be discovered),
as we will discuss later.

Among the results in Table \ref{tab:partners} it is interesting to notice
the strange-isoscalar $P_c^{\Lambda}$ partners of the three LHCb pentaquarks,
which were predicted (together with the pentaquarks) nearly
a decade ago~\cite{Wu:2010jy,Wu:2010vk}.
This prediction has been recently updated in Ref.~\cite{Xiao:2019gjd},
which uses a contact-range theory where the couplings are saturated
by vector{-}meson exchange and the regularization
is set as to reproduce the $P_c(4312)$ pentaquark.
The prediction of Ref.~\cite{Xiao:2019gjd} for the mass of
the $\bar{D} \Xi_c'$ molecule is $4436.7\,{\rm MeV}$,
which happens to be pretty close to ours
(check Table \ref{tab:partners}).
Refs.~\cite{Gutsche:2019mkg,Wang:2019nvm} have also made
a series of molecular pentaquark predictions which closely match ours.

{
  On the experimental side it is worth mentioning that a $P_c^{\Lambda}$
  pentaquark --- the $P_{cs}(4459)$ --- has been observed by
  the LHCb collaboration~\cite{LHCb:2020jpq},
  but owing to its mass it is probably a $\bar{D}^* \Xi_c$
  molecule~\cite{Chen:2020uif,Peng:2020hql,Chen:2020kco,Liu:2020hcv}.
  As such it involves a $\bar{3}$ charmed baryon ($\Lambda_c$, $\Xi_c$)
  instead of a sextet one ($\Sigma_c$, $\Xi_c'$, $\Omega_c$ and
  their excited states), which means that this pentaquark
  is not expected to be one of the SU(3)-flavor partners
  of the $P_c(4312)$, $P_c(4440)$ and $P_c(4457)$ that
  we predict here.
  Nonetheless, the $P_{cs}(4459)$ will prove useful as a phenomenological
  cross-check of the size of SU(3)-flavor violations,
  as we will argue later.
  Regarding the possible five-flavor partners of the $P_{cs}(4459)$,
  there is a recent exploration in Ref.~\cite{Shen:2022rpn}.
}

\section{Uncertainties}

We are predicting the molecular pentaquarks within a contact-range EFT,
which entails that they are amenable to systematic error estimations.
A conventional way to estimate these theoretical errors is
to vary the predictions within a sensible cutoff window
(which is what we have done for the five-flavor pentaquarks
in Eqs.~(\ref{eq:P5-a}-\ref{eq:P5-d})).
Here the cutoff floats from $0.5\,$ to $1\,{\rm GeV}$,
which can be either identified
with the mass of the vector mesons or with the momenta
at which the internal structure of the hadrons starts to be resolved.
For the $c\bar{c}$ family of pentaquarks this translates into a systematic
error of less than $1\,{\rm MeV}$, which explains why the predictions
of other theoretical works~\cite{Xiao:2019gjd,Gutsche:2019mkg,Wang:2019nvm}
are basically identical to ours.
Yet this uncertainty is calculated under the assumption that SU(3)-flavor
symmetry is perfectly preserved, which is not the case.
Violations of SU(3)-flavor symmetry relations are usually of
the order of $20\%$, as estimated from the difference between
the pion and kaon weak decay constants ($f_{\pi} \simeq 130\,{\rm MeV}$
and $f_{K} \simeq 160\,{\rm MeV}$).
From this, within the EFT we are using we can be easily take into account
the SU(3)-flavor symmetry breaking effects by randomly varying
the $C^O_i$ couplings by $20\%$ around
their central values.
For $\Lambda = 0.75\,{\rm GeV}$, this translates into an uncertainty of
$2-15\,{\rm MeV}$ depending on the specific $c\bar{c}$ pentaquark,
where the largest uncertainties correspond to the states
with the largest binding energies.

For the $c \bar{b}$, $\bar{c} b$ and $b \bar{b}$ molecular pentaquarks
the situation is different owing to the considerably larger cutoff
dependence (about $5$, $10$ and $20-30\,{\rm MeV}$ respectively),
which we will discuss in the next paragraph.
The SU(3)-flavor uncertainties in these cases will be $10-20$ and
$15-25\,{\rm MeV}$ for the $c \bar{b}$/$\bar{c} b$ and $b \bar{b}$
cases, respectively.
That is, while for the $c \bar{c}$, $c \bar{b}$, $\bar{c} b$ the uncertainties
are dominated by flavor symmetry breaking effects,
for the $b \bar{b}$ pentaquarks cutoff variation
tends to be the largest source of uncertainty.

{
However, the application of SU(3)-flavor symmetry remains theoretical
in the sense that we do not really have a clear molecular example
from where we can determine how well this symmetry works
at the quantitative level.
Two qualitative examples are already known:
\begin{itemize}
\item[(i)] The $Z_c(3900)$~\cite{Ablikim:2013mio} and
  $Z_{cs}(3895)$~\cite{Ablikim:2020hsk} ($Z_c$ and $Z_{cs}$ from now on),
  which have been theorized to be $I=1$
  $D^*\bar{D}$~\cite{Wang:2013cya,Guo:2013sya,Albaladejo:2015lob} and
  $I=\tfrac{1}{2}$ $D_s^* \bar{D} - D_s \bar{D}^*$~\cite{Yang:2020nrt,Sun:2020hjw} molecules, respectively.
\item[(ii)] The $P_{cs}(4459)$ pentaquark~\cite{LHCb:2020jpq},
  which has been theorized to be an $I=0$ $\bar{D}^* \Xi_c$
  bound state~\cite{Chen:2020uif,Peng:2020hql,Chen:2020kco,Liu:2020hcv}.
\end{itemize}
In the first case, the SU(3) decomposition of heavy meson-antimeson states is
$3 \otimes \bar{3} = 1 \oplus 8$, i.e. a singlet and an octet representation,
where the $Z_c$ and $Z_{cs}$ both belong to the octet and thus their
potential is expected to be the same~\cite{HidalgoDuque:2012pq,Yang:2020nrt}.
But it happens that the masses of the $Z_c$ and $Z_{cs}$ resonances are above
their corresponding meson-antimeson thresholds, which means that they are not
necessarily bound states but more probably resonances (or even virtual
states if we take into account that their Breit-Wigner masses might
not correspond to their physical masses).
If this happens to be the case, they will require a different contact-range EFT
description than the one we employ here for the pentaquarks (or the direct
extraction of the couplings from the data instead of the masses,
as done in Refs.~\cite{Albaladejo:2015lob,Yang:2020nrt}),
which renders it difficult to make direct comparisons
between the $Z_c$'s and the $P_c$'s.
}

{
  In the second case, as pointed out previously, the $\Xi_c$ charmed baryon is
  a flavor antitriplet and the $\bar{D}^* \Xi_c$ system will essentially
  belong to a different and independent representation of SU(3).
  That is, the $\bar{D}^* \Xi_c$ potential can be described with a new coupling
  constant $D(\Lambda)$, i.e.
  \begin{eqnarray}
    \langle p' | V | p \rangle = D(\Lambda)\,
    f(\frac{p}{\Lambda})\,f(\frac{p'}{\Lambda}) \, ,
  \end{eqnarray}
  the value of which is in principle unrelated to the $C^O_i(\Lambda)$ couplings
  we have used to reproduce the three $P_c$ pentaquarks.
  However, phenomenological models based on vector-meson exchanges predict
  that $D = C^O_1$~\cite{Wu:2010jy,Wu:2010vk},
  i.e. the $I=0$ $\bar{D}^* \Xi_c$ and $I=\tfrac{1}{2}$
  $\bar{D} \Sigma_c$ potentials are expected to be similar.
  Concrete calculations with the same type of EFT, regulator and cutoff range
  we have used for the $P_{c1}$, $P_{c2}$ and $P_{c3}$ yield $D = 1.17\,C^O_1$
  when calibrating $D(\Lambda)$ to the $P_{cs}(4459)$ mass,
  showing a $17\%$ discrepancy from $D = C^O_1$.
  The more complete analysis of Ref.~\cite{Peng:2020hql} (which includes a
  series of effects not considered here, like coupled channel dynamics or
  the double-peak solution considered in the experimental analysis of
  Ref.~\cite{LHCb:2020jpq}) provides a compatible figure of
  $D = (0.90-1.11)\,C^O_1$, which deviates a merely $10\%$
  away from the phenomenological relation $D = C^O_1$.
  The previous numbers are well within the $20\%$ SU(3) uncertainty
  estimated from the $f_{\pi}$ and $f_{K}$ difference. This is despite
  the fact that the $D = C^O_1$ relation is based on phenomenology,
  from which further uncertainties (beyond SU(3) symmetry breaking)
  should be expected.
}

Regarding HFS, as already pointed out, its application beyond the $c\bar{c}$
sector has a serious limitation in terms of model dependence
within the contact-range EFT framework.
The cutoff dependence of the predictions becomes larger
as the reduced mass of the system is increased,
from merely 1{ }${\rm MeV}$ at most in the hidden-charm sector
to a couple of tens of ${\rm MeV}$ in the hidden-bottom sector.
This limitation was already pointed out in Ref.~\cite{Baru:2018qkb},
where here we merely confirm the impossibility of
making model independent predictions with HFS.
Yet we notice that there is {\it systematicity} in this model dependence,
as {increasing the cutoff $\Lambda$} invariably leans towards more binding.
This is important, as it implies that the conclusion that
the $c \bar{b}$, $\bar{c} b$ and $b \bar{b}$ molecular pentaquarks bind
is indeed model independent, with the model dependence
limited to how much they bind.
In fact it can be shown that for two-body molecular systems
where the potential respects HFS (i.e. the potential is
independent of the heavy-quark mass), the binding energy $B_2$
increases monotonically with the reduced mass $\mu$,
$\partial B_2 / \partial \mu \geqslant 0$
(check Appendix \ref{app:binding-mass}
for further details).
That is, though the specific masses of the $c \bar{b}$, $\bar{c} b$
and $b \bar{b}$ pentaquarks are model dependent to a certain extent,
the fact that these systems bind is a model independent
outcome of the calculations.

\section{Including heavy-quark spin symmetry}

%
Previously we have made the simplifying assumption that the potentials
binding the $P_{c1}$, $P_{c2}$ and $P_{c3}$ pentaquarks are unrelated.
However, HQSS connects the potentials of these three configurations and
\mpv{allows for a common description of the $\bar{P} \Sigma_Q$,
  $\bar{P}^* \Sigma_Q$ and $\bar{P}^* \Sigma_Q^*$
  molecules~\cite{Xiao:2013yca,Yamaguchi:2016ote,Yamaguchi:2017zmn,Shimizu:2018ran} (where here we will concentrate on the consequences of HQSS
  for the type of contact-range EFTs we are using).}
The disadvantage though is that we do not know which of the $P_{c2}$ and $P_{c3}$
pentaquarks corresponds to the $J=\tfrac{1}{2}$ and $\tfrac{3}{2}$
$\bar{D}^* \Sigma_c$ configurations.
As a consequence there are two possible set of predictions for
the $\bar{P}^{(*)} \Sigma_Q^{(*)}$ family of molecules,
depending on which spin identification we propose
for the $P_{c2}$ and $P_{c3}$ pentaquarks.

HQSS indicates that the $| \bar{Q} q \rangle $ and $| Q q q \rangle$
family of heavy hadrons are related by means of rotations of
the spin of the heavy quark.
Indeed, we can group the ground and excited states of a heavy hadron
in a single superfield, which for the S-wave heavy mesons and baryons
are defined as
\begin{eqnarray}
  H &=& \frac{1}{\sqrt{2}}\,
  \left[ P + \vec{\sigma} \cdot \vec{P}^* \right] \, , \\
  \vec{S} &=& \frac{1}{\sqrt{3}}\,\vec{\sigma}\,B_6 + \vec{B}^*_6 \, ,
\end{eqnarray}
where for simplicity we are ignoring the SU(3)-flavor indices and with
$P$, $P^*$ the $J=0,1$ heavy mesons, $B_6$, $B^*_6$
the $J=\tfrac{1}{2}$, $\tfrac{3}{2}$ heavy baryons
and $\vec{\sigma}$ the Pauli matrices.
With the previous definitions, the lowest-order contact-range Lagrangian
describing molecular pentaquarks reads~\cite{Liu:2018zzu}
\begin{eqnarray}
  \mathcal{L}_{\rm contact} &=&
  C_a\,{\rm Tr}[H^{\dagger} H]\,\vec{S}^{\dagger}\cdot \vec{S} \nonumber \\
  &+& C_b\,\sum_{i=1}^3\,{\rm Tr}[H^{\dagger} \sigma_i H]\,
  \vec{S}^{\dagger} \cdot (J_i \vec{S}) \, , 
\end{eqnarray}
where $J_i$ are the $i=1,2,3$ spin-1 matrices.
The terms proportional to the couplings $C_a$ and $C_b$ correspond to
central and spin-spin contact-range interactions.
Thus, the practical implication of the HQSS version of the contact-range
Lagrangian is that the $C^O_i$ couplings we previously defined
in Eq.~(\ref{eq:L-contact}) can be decomposed
in central and spin-spin components:
\begin{eqnarray}
  C_i^O \to C_a^O + \lambda_i\,C_b^O  \, ,
\end{eqnarray}
where the explicit decomposition for the three known molecular pentaquark
candidates is
\begin{eqnarray}
  V_C(\bar{P} \Sigma_Q) &=& C_a^O \, , \\
  V_C(\bar{P}^* \Sigma_Q, J=\tfrac{1}{2}) &=& C_a^O - \frac{4}{3}\,C_b^O\, , \\
  V_C(\bar{P}^* \Sigma_Q, J=\tfrac{3}{2}) &=& C_a^O + \frac{2}{3}\,C_b^O\, ,
\end{eqnarray}
while for the four potentially unobserved configurations we will have
\begin{eqnarray}
  V_C(\bar{P} \Sigma_Q^*) &=& C_a^O \, , \\
  V_C(\bar{P}^* \Sigma_Q^*, J=\tfrac{1}{2}) &=& C_a^O - \frac{5}{3}\,C_b^O\, , \\
  V_C(\bar{P}^* \Sigma_Q^*, J=\tfrac{3}{2}) &=& C_a^O - \frac{2}{3}\,C_b^O\, , \\
  V_C(\bar{P}^* \Sigma_Q^*, J=\tfrac{5}{2}) &=& C_a^O +C_b^O\, .
\end{eqnarray}

Now, for the $P_{c1}$ pentaquark the identification of
its particle and spin channel is trivial:
$J=\tfrac{1}{2}$ $\bar{D} \Sigma_c$.
Meanwhile this is not the case for the $P_{c2}$ and $P_{c3}$ pentaquarks:
both are expected to be $\bar{D}^* \Sigma_c$ molecules, but what is not
clear is which one is the spin $J=\tfrac{1}{2}$ and $\tfrac{3}{2}$
state, as their spins have not been experimentally determined yet.
Thus there are two possibilities:
\begin{itemize}
\item[(i)] that the $P_{c2}$ and $P_{c3}$ pentaquarks
are $J=\tfrac{1}{2}$ and $\tfrac{3}{2}$ states, respectively, thus following
the standard pattern of mass increasing with spin, which we will call
scenario A, and
\item[(ii)] the opposite pattern, mass decreasing with spin,
  is scenario B.
\end{itemize}
These scenarios have been named following the convention found
in Ref.~\cite{Liu:2019tjn}.
Different theoretical works prefer scenario A~\cite{Chen:2019asm,Wang:2019ato},
scenario B~\cite{Yamaguchi:2019seo,Liu:2019zvb,Du:2019pij,Yalikun:2021bfm},
do not find a strong preference~\cite{Liu:2019tjn,PavonValderrama:2019nbk}
or explore alternative possibilities~\cite{Burns:2019iih,Burns:2021jlu}.
\mpv{
  Scenario A has recently been explained as a consequence of the short-range
  interaction of the light-quarks within the heavy antimeson and
  heavy baryon composing the pentaquarks~\cite{Chen:2021cfl}.
  Scenario B appeared before the discovery of the pentaquark trio, for instance
  in Ref.~\cite{Yamaguchi:2016ote}, and has received explanations both
  in terms of pion~\cite{Karliner:2015ina} and vector
  meson exchanges~\cite{Peng:2020xrf}.
}

Here, we will calibrate the $C_a^O$ and $C_b^O$ couplings to the masses of
the $P_{c1}$ and $P_{c3}$ pentaquarks in scenarios A and B, leading to
\begin{eqnarray}
  C_a^O &=& -1.17\,(-(0.78-2.16))\,{\rm fm}^2 \,\,\mbox{(A)} \, , \\
  C_b^O &=& +0.21\,(+(0.11-0.54))\,{\rm fm}^2 \,\,\mbox{(A)} \, , \\
  \nonumber \\
  C_a^O &=& -1.30\,(-(0.85-2.52))\,{\rm fm}^2 \,\,\mbox{(B)} \, , \\
  C_b^O &=& -0.21\,(-(0.11-0.54))\,{\rm fm}^2 \,\,\mbox{(B)} \, , 
\end{eqnarray}
depending on the scenario, where the intervals in parentheses refer to
the cutoff variation (i.e. $(0.5-1.0)\,{\rm GeV}$).
From this we can calculate the complete spectrum of the $\bar{D}^* \Sigma_c$,
$\bar{D}^* \Sigma_c^*$ and their SU(3)- and heavy-flavor
counterparts, where we show the results in Tables
\ref{tab:HQS-partners-ccbar-bcbar} ($c\bar{c}$ and $b \bar{c}$ sectors) and
\ref{tab:HQS-partners-cbbar-bbbar} ($c\bar{b}$ and $b \bar{b}$ sectors).
We find that most pentaquark configurations (112 in total)
bind within theoretical uncertainties
(which are computed as before).

\begin{table*}[!ttt]
\begin{tabular}{|ccccccc|ccccccc|}
\hline \hline
Molecule & $I$ & $S$ & $B_P$ & $M_P$ & $J$ & Scenario
&
Molecule & $I$ & $S$ & $B_P$ & $M_P$ & $J$ & Scenario
\\
  \hline
  $\bar{D} \Sigma_c^*$ & $\tfrac{1}{2}$ & \phantom{+}$0$ &
  $8.4^{+0.5}_{-0.4}$ & $4376.9^{+0.4}_{-0.5}$
  & $\frac{3}{2}$ & A
  &
  $\bar{D} \Sigma_c^*$ & $\tfrac{1}{2}$ & \phantom{+}$0$ &
  $14.0^{+0.6}_{-0.6}$ & $4371.4^{+0.6}_{-0.6}$
  & $\frac{3}{2}$ & B
  \\
  $\bar{D}^* \Sigma_c^*$ & $\tfrac{1}{2}$ & \phantom{+}$0$ &
  $25.9^{+0.3}_{-0.4}$ & $4500.7^{+0.4}_{-0.3}$
  & $\frac{1}{2}$ & A
  &
  $\bar{D}^* \Sigma_c^*$ & $\tfrac{1}{2}$ & \phantom{+}$0$ &
  $3.2^{+0.2}_{-0.2}$ & $4523.5^{+0.2}_{-0.2}$
  & $\frac{1}{2}$ & B
  \\
  $\bar{D}^* \Sigma_c^*$ & $\tfrac{1}{2}$ & \phantom{+}$0$ &
  $15.8^{+0.1}_{-0.0}$ & $4510.9^{+0.0}_{-0.1}$
  & $\frac{3}{2}$ & A
  &
  $\bar{D}^* \Sigma_c^*$ & $\tfrac{1}{2}$ & \phantom{+}$0$ &
  $9.9^{+0.1}_{-0.0}$ & $4516.8^{+0.0}_{-0.0}$
  & $\frac{3}{2}$ & B
  \\
  $\bar{D}^* \Sigma_c^*$ & $\tfrac{1}{2}$ & \phantom{+}$0$ &
  $3.2^{+0.1}_{-0.2}$ & $4523.5^{+0.2}_{-0.1}$
  & $\frac{5}{2}$ & A
  &
  $\bar{D}^* \Sigma_c^*$ & $\tfrac{1}{2}$ & \phantom{+}$0$ &
  $25.9^{+0.3}_{-0.4}$ & $4500.7^{+0.4}_{-0.3}$
  & $\frac{5}{2}$ & B
  \\
  \hline
  $\bar{D} \Xi_c^*$ & $0$ & $-1$ &
  $9.2^{+10.1}_{-7.1}$ & $4503.7^{+7.1}_{-10.1}$ & $\frac{3}{2}$ & A
  &
  $\bar{D} \Xi_c^*$ & $0$ & $-1$ &
  $15^{+13}_{-10}$ & $4498^{+10}_{-13}$ & $\frac{3}{2}$ & B
  \\
  $\bar{D}^* \Xi_c^*$ & $0$ & $-1$ &
  $27^{+17}_{-15}$ & $4627^{+15}_{-17}$ & $\frac{1}{2}$ & A
  &
  $\bar{D}^* \Xi_c^*$ & $0$ & $-1$ &
  $3.6^{+6.5}_{-3.5}$ & $4650.5^{+3.5}_{-6.5}$ & $\frac{1}{2}$ & B
  \\
  $\bar{D}^* \Xi_c^*$ & $0$ & $-1$ &
  $17^{+13}_{-11}$ & $4638^{+11}_{-13}$ & $\frac{3}{2}$ & A
  &
  $\bar{D}^* \Xi_c^*$ & $0$ & $-1$ &
  $10.7^{+6.5}_{-7.8}$ & $4643.5^{+7.8}_{-6.5}$ & $\frac{3}{2}$ & B
  \\
  $\bar{D}^* \Xi_c^*$ & $0$ & $-1$ &
  $3.5^{+6.5}_{-3.4}$ & $4650.5^{+3.4}_{-6.5}$ & $\frac{5}{2}$ & A
  &
  $\bar{D}^* \Xi_c^*$ & $0$ & $-1$ &
  $27^{+17}_{-15}$ & $4627^{+15}_{-17}$ & $\frac{5}{2}$ & B
  \\
  \hline
  $\bar{D} \Xi_c^*-\bar{D}_s \Sigma_c^*$ & $1$ & $-1$ &
  $4.5^{+8.7}_{-4.4}$ & $4481.9^{+4.4}_{-8.7}$ &
  $\frac{3}{2}$ & A
  &
  $\bar{D} \Xi_c^*-\bar{D}_s \Sigma_c^*$  & $1$ & $-1$ &
  $9.6^{+12.0}_{-8.0}$ & $4477.0^{+12.0}_{-8.0}$ & $\frac{3}{2}$ & B
  \\
  $\bar{D}^* \Xi_c^*-\bar{D}_s^* \Sigma_c^*$ & $1$ & $-1$ &
  $21^{+17}_{-14}$ & $4609^{+14}_{-17}$ & $\frac{1}{2}$ & A
  &
  $\bar{D}^* \Xi_c^*-\bar{D}_s^* \Sigma_c^*$ & $1$ & $-1$ &
  $0.8^{+4.8}_{-0.8}$ & $4629.5^{+0.8}_{-4.8}$ & $\frac{1}{2}$ & B
  \\
  $\bar{D}^* \Xi_c^*-\bar{D}_s^* \Sigma_c^*$ & $1$ & $-1$ &
  $11.4^{+12.7}_{-9.0}$ & $4618.9^{+9.0}_{-12.7}$ & $\frac{3}{2}$ & A
  &
  $\bar{D}^* \Xi_c^*-\bar{D}_s^* \Sigma_c^*$ & $1$ & $-1$ &
  $6.1^{+9.7}_{-5.6}$ & $4624.3^{+5.6}_{-9.7}$ & $\frac{3}{2}$ & B
  \\
  $\bar{D}^* \Xi_c^*-\bar{D}_s^* \Sigma_c^*$ & $1$ & $-1$ &
  $0.8^{+4.8}_{-0.8}$ & $4629.5^{+0.8}_{-4.8}$ & $\frac{5}{2}$ & A
  &
  $\bar{D}^* \Xi_c^*-\bar{D}_s^* \Sigma_c^*$ & $1$ & $-1$ &
  $21^{+17}_{-14}$ & $4609^{+14}_{-17}$ & $\frac{5}{2}$ & B
  \\
  \hline
  $\bar{D} \Omega_c^*-\bar{D}_s \Xi_c^*$ & $\tfrac{1}{2}$ & $-2$ &
  $1.4^{+8.4}_{-1.4}$ & $4612.6^{+1.4}_{-8.4}$ & $\frac{1}{2}$ & A
  &
  $\bar{D} \Omega_c^*- \bar{D}_s \Xi_c^*$ & $\tfrac{1}{2}$ & $-2$
  & $5.7^{+12.0}_{-5.7}$ & $4608.2^{+5.7}_{-12.0}$ & $\frac{1}{2}$ & B
  \\
  $\bar{D}^* \Omega_c^*-\bar{D}_s^* \Xi_c^*$ & $\tfrac{1}{2}$ & $-2$
  & $18^{+17}_{-14}$ & $4739^{+14}_{-17}$ & $\frac{1}{2}$ & A
  &
  $\bar{D}^* \Omega_c^*-\bar{D}_s^* \Xi_c^*$ & $\tfrac{1}{2}$ & $-2$
  & $-{(2.7)}$ & $-{(4755.1)}$ & $\frac{1}{2}$ & B
  \\
  $\bar{D}^* \Omega_c^*-\bar{D}_s^* \Xi_c^*$ & $\tfrac{1}{2}$ & $-2$
  & $8.4^{+12.9}_{-8.2}$ & $4749.4^{+8.2}_{-12.9}$ & $\frac{3}{2}$ & A
  &
  $\bar{D}^* \Omega_c^*-\bar{D}_s^* \Xi_c^*$ & $\tfrac{1}{2}$ & $-2$
  & $3.1^{+9.7}_{-3.1}$ & $4754.7^{+3.1}_{-9.7}$ & $\frac{3}{2}$ & B
  \\
  $\bar{D}^* \Omega_c^*-\bar{D}_s^* \Xi_c^*$ & $\tfrac{1}{2}$ & $-2$
  & $-{(2.7)}$ & $-{(4755.1)}$ & $\frac{5}{2}$ & A
  &
  $\bar{D}^* \Omega_c^*-\bar{D}_s^* \Xi_c^*$ & $\tfrac{1}{2}$ & $-2$
  & $18^{+17}_{-14}$ & $4739^{+14}_{-17}$ & $\frac{5}{2}$ & B
  \\
  \hline \hline
  $\bar{D} \Sigma_b^*$ & $\tfrac{1}{2}$ & \phantom{+}$0$
  & $19.1^{+4.3}_{-3.7}$ & $7680.7^{+3.7}_{-4.3}$ & $\frac{3}{2}$ & A
  &
  $\bar{D} \Sigma_b^*$ & $\tfrac{1}{2}$ & \phantom{+}$0$
  & $26.6^{+5.0}_{-3.6}$ & $7673.2^{+3.6}_{-5.0}$  & $\frac{3}{2}$ & B
  \\
  $\bar{D}^* \Sigma_b^*$ & $\tfrac{1}{2}$ & \phantom{+}$0$
  & $41.9^{+7.8}_{-7.2}$ & $7799.2^{+7.2}_{-7.8}$ & $\frac{1}{2}$ & A
  &
  $\bar{D}^* \Sigma_b^*$ & $\tfrac{1}{2}$ & \phantom{+}$0$
  & $11.1^{+4.3}_{-3.6}$ & $7830.0^{+3.6}_{-4.3}$ & $\frac{1}{2}$ & B
  \\
  $\bar{D}^* \Sigma_b^*$ & $\tfrac{1}{2}$ & \phantom{+}$0$
  & $29.2^{+6.3}_{-5.6}$ & $7811.9^{+5.6}_{-6.3}$ & $\frac{3}{2}$ & A
  &
  $\bar{D}^* \Sigma_b^*$ & $\tfrac{1}{2}$ & \phantom{+}$0$
  & $21.4^{+5.5}_{-4.7}$ & $7819.7^{+4.7}_{-5.5}$ & $\frac{3}{2}$ & B
  \\
  $\bar{D}^* \Sigma_b^*$ & $\tfrac{1}{2}$ & \phantom{+}$0$
  & $11.1^{+4.3}_{-3.6}$ & $7830.0^{+3.6}_{-4.3}$ & $\frac{5}{2}$ & A
  &
  $\bar{D}^* \Sigma_b^*$ & $\tfrac{1}{2}$ & \phantom{+}$0$
  & $41.9^{+7.8}_{-7.1}$ & $7799.2^{+7.1}_{-7.8}$ & $\frac{5}{2}$ & B
  \\
  \hline
  $\bar{D} \Xi_b^*$ & $0$ & $-1$ &
  $22^{+18}_{-10}$ & $7800^{+10}_{-17}$ & $\frac{3}{2}$ & A
  &
  $\bar{D} \Xi_b^*$ & $0$ & $-1$
  & $29^{+21}_{-12}$ & $7792^{+12}_{-21}$ & $\frac{3}{2}$ & B
  \\
  $\bar{D}^* \Xi_b^*$ & $0$ & $-1$
  & $42^{+22}_{-20}$ & $7920^{+20}_{-22}$ & $\frac{1}{2}$ & A
  &
  $\bar{D}^* \Xi_b^*$ & $0$ & $-1$ &
  $11.3^{+10.6}_{-8.3}$ & $7951.1^{+8.3}_{-10.6}$ & $\frac{1}{2}$ & B
  \\
  $\bar{D}^* \Xi_b^*$ & $0$ & $-1$ &
  $29^{+17}_{-15}$ & $7933^{+15}_{-17}$ & $\frac{3}{2}$ & A
  &
  $\bar{D}^* \Xi_b^*$ & $0$ & $-1$ &
  $22^{+15}_{-13}$ & $7941^{+13}_{-15}$ & $\frac{3}{2}$ & B
  \\
  $\bar{D}^* \Xi_b^*$ & $0$ & $-1$ &
  $11.3^{+10.5}_{-8.3}$ & $7951.1^{+8.3}_{-10.5}$ & $\frac{5}{2}$ & A
  &
  $\bar{D}^* \Xi_b^*$ & $0$ & $-1$ &
  $42^{+22}_{-20}$ & $7920^{+20}_{-22}$ & $\frac{5}{2}$ & B
  \\
  \hline
  $\bar{D} \Xi_b^*-\bar{D}_s \Sigma_b^*$ & $1$ & $-1$
  & $15^{+14}_{-11}$ & $7786^{+11}_{-14}$ &
  $\frac{3}{2}$ & A
  &
  $\bar{D} \Xi_b^*-\bar{D}_s \Sigma_b^*$  & $1$ & $-1$
  & $23^{+17}_{-14}$ & $7778^{+14}_{-17}$ & $\frac{3}{2}$ & B
  \\
  $\bar{D}^* \Xi_b^*-\bar{D}_s^* \Sigma_b^*$ & $1$ & $-1$
  & $38^{+22}_{-20}$ & $7906^{+20}_{-22}$ & $\frac{1}{2}$ & A
  &
  $\bar{D}^* \Xi_b^*-\bar{D}_s^* \Sigma_b^*$ & $1$ & $-1$
  & $8.0^{+10.2}_{-7.2}$ & $7936.7^{+7.2}_{-10.2}$ & $\frac{1}{2}$ & B
  \\
  $\bar{D}^* \Xi_b^*-\bar{D}_s^* \Sigma_b^*$ & $1$ & $-1$
  & $28^{+18}_{-15}$ & $7919^{+15}_{-18}$ & $\frac{3}{2}$ & A
  &
  $\bar{D}^* \Xi_b^*-\bar{D}_s^* \Sigma_b^*$ & $1$ & $-1$
  & $18^{+15}_{-12}$ & $7927^{+12}_{-15}$ & $\frac{3}{2}$ & B
  \\
  $\bar{D}^* \Xi_b^*-\bar{D}_s^* \Sigma_b^*$ & $1$ & $-1$
  & $8.0^{+10.2}_{-7.2}$ & $7936.7^{+7.2}_{-10.2}$ & $\frac{5}{2}$ & A
  &
  $\bar{D}^* \Xi_b^*-\bar{D}_s^* \Sigma_b^*$ & $1$ & $-1$
  & $38^{+22}_{-20}$ & $7906^{+20}_{-22}$ & $\frac{5}{2}$ & B
  \\
  \hline
  $\bar{D} \Omega_b^*-\bar{D}_s \Xi_b^*$ & $\tfrac{1}{2}$ & $-2$ &
  $13^{+14}_{-11}$ & $7909^{+11}_{-14}$ & $\frac{3}{2}$ & A
  &
  $\bar{D} \Omega_b^*- \bar{D}_s \Xi_b^*$ & $\tfrac{1}{2}$ & $-2$
  & $21^{+17}_{-14}$ & $7901^{+14}_{-17}$ & $\frac{3}{2}$ & B
  \\
  $\bar{D}^* \Omega_b^*-\bar{D}_s^* \Xi_b^*$ & $\tfrac{1}{2}$ & $-2$
  & $38^{+22}_{-20}$ & $8029^{+20}_{-22}$ & $\frac{1}{2}$ & A
  &
  $\bar{D}^* \Omega_b^*-\bar{D}_s^* \Xi_b^*$ & $\tfrac{1}{2}$ & $-2$
  & $6.8^{+10.4}_{-7.3}$ & $8059.2^{+7.3}_{-10.4}$ & $\frac{1}{2}$ & B
  \\
  $\bar{D}^* \Omega_b^*-\bar{D}_s^* \Xi_b^*$ & $\tfrac{1}{2}$ & $-2$
  & $25^{+18}_{-15}$ & $8041^{+15}_{-18}$ & $\frac{3}{2}$ & A
  &
  $\bar{D}^* \Omega_b^*-\bar{D}_s^* \Xi_b^*$ & $\tfrac{1}{2}$ & $-2$
  & $17^{+15}_{-12}$ & $8049^{+12}_{-15}$ & $\frac{3}{2}$ & B
  \\
  $\bar{D}^* \Omega_b^*-\bar{D}_s^* \Xi_b^*$ & $\tfrac{3}{2}$ & $-2$
  & $6.8^{+10.4}_{-7.3}$ & $8059.2^{+7.3}_{-10.4}$ & $\frac{5}{2}$ & A
  &
  $\bar{D}^* \Omega_b^*-\bar{D}_s^* \Xi_b^*$ & $\tfrac{5}{2}$ & $-2$
  & $38^{+22}_{-20}$ & $8029^{+20}_{-22}$ & $\frac{5}{2}$ & B
  \\
  \hline \hline
\end{tabular}
\caption{
  The heavy-quark spin, heavy-flavor and light-flavor symmetry partners of
  the $P_c(4312)$, $P_c(4440)$ and $P_c(4457)$ pentaquarks,
  where in this table we consider the configurations
  with heavy-quark content $c\bar{c}$ and $b\bar{c}$.
  The predictions depend on which are the assumptions made for the spin of
  the $P_c(4440)$ and $P_c(4457)$ pentaquarks: scenario A refers to the
  $P_c(4440)$ and $P_c(4457)$ having spin $J=\tfrac{1}{2}$ and
  $\tfrac{3}{2}$, while scenario B considers
  the opposite identification.
  The columns ``Molecule'', $I$, $S$, $B_P$ and $M_P$ have the same meaning
  as in Table~\ref{tab:partners}, while $J$ refers to the spin of the
  molecular configuration and ``Scenario'' to the two aforementioned
  possibilities (A \& B).
  In the coupled-channel cases, the binding energy is calculated
  relative to the channel with the lowest mass.
  The calculations use the contact-range EFT of Eq.~(\ref{eq:contact}) and
  a Gaussian regulator with a cutoff $\Lambda = 0.75\,{\rm GeV}$.
  The uncertainties are obtained from two sources (and then summed
  in quadrature): the error coming from varying the cutoff
  in the $(0.5-1.0)\,{\rm GeV}$ window and an expected
  violation of SU(3)-flavor symmetry of $20\%$
  in the contact-range couplings
  (this later error only applies to configurations containing strangeness).
  The notation $- {(B_P/M_P)}$ indicates a configuration that does not bind
  for the central estimation of the parameters, but could have
  binding energy $B_P$ / mass $M_P$ within uncertainties.
  For the mass of the $\Omega_b^*$ (which has not been experimentally
  observed yet), we simply assume $m(\Omega_b^*) - m(\Omega_b)
  \simeq m(\Xi_b^*) - m(\Xi_b') \simeq
  m(\Sigma_b^*) - m(\Sigma_b') \simeq 20\,{\rm MeV}$;
  the effect of the $\Omega_b^*$ mass on the predictions of
  the $\bar{P}^* \Omega_b^*$-$\bar{P}_s^* \Xi_b^*$ pentaquarks is minimal
  though because the lowest mass threshold corresponds to
  the $\bar{P}_s^* \Xi_b^*$ two-hadron system.
}
\label{tab:HQS-partners-ccbar-bcbar}
\end{table*}

\begin{table*}[!ttt]
\begin{tabular}{|ccccccc|ccccccc|}
\hline \hline
Molecule & $I$ & $S$ & $B_P$ & $M_P$ & $J$ & Scenario
&
Molecule & $I$ & $S$ & $B_P$ & $M_P$ & $J$ & Scenario
\\
  \hline
  $B \Sigma_c^*$ & $\tfrac{1}{2}$ & \phantom{+}$0$
  & $27.0^{+8.6}_{-7.3}$ & $7770.6^{+7.3}_{-8.6}$ & $\frac{3}{2}$ & A
  &
  $B \Sigma_c^*$ & $\tfrac{1}{2}$ & \phantom{+}$0$
  & $35.5^{+9.8}_{-8.4}$ & $7762.2^{+8.4}_{-9.8}$ & $\frac{3}{2}$ & B
  \\
  $B^* \Sigma_c^*$ & $\tfrac{1}{2}$ & \phantom{+}$0$
  & $49^{+12}_{-10}$ & $7794^{+10}_{-12}$ & $\frac{1}{2}$ & A
  &
  $B^* \Sigma_c^*$ & $\tfrac{1}{2}$ & \phantom{+}$0$
  & $15.6^{+7.0}_{-5.8}$ & $7827.3^{+5.8}_{-7.0}$ & $\frac{1}{2}$ & B
  \\
  $B^* \Sigma_c^*$ & $\tfrac{1}{2}$ & \phantom{+}$0$
  & $35.6^{+9.8}_{-8.5}$ & $7807.3^{+8.5}_{-9.8}$ & $\frac{3}{2}$ & A
  &
  $B^* \Sigma_c^*$ & $\tfrac{1}{2}$ & \phantom{+}$0$
  & $27.1^{+8.7}_{-7.3}$ & $7815.7^{+7.3}_{-8.7}$ & $\frac{3}{2}$ & B
  \\
  $B^* \Sigma_c^*$ & $\tfrac{1}{2}$ & \phantom{+}$0$
  & $15.6^{+7.0}_{-5.8}$ & $7827.3^{+5.8}_{-7.0}$ & $\frac{5}{2}$ & A
  &
  $B^* \Sigma_c^*$ & $\tfrac{1}{2}$ & \phantom{+}$0$
  & $49^{+12}_{-10}$ & $7794^{+10}_{-12}$ & $\frac{5}{2}$ & B
  \\
  \hline
  $B \Xi_c^*$ & $0$ & $-1$
  & $28^{+18}_{-15}$ & $7897^{+15}_{-18}$ & $\frac{3}{2}$ & A
  &
  $B \Xi_c^*$ & $0$ & $-1$
  & $37^{+21}_{-18}$ & $7888^{+18}_{-21}$ & $\frac{3}{2}$ & B
  \\
  $B^* \Xi_c^*$ & $0$ & $-1$
  & $51^{+25}_{-23}$ & $7919^{+23}_{-25}$ & $\frac{1}{2}$ & A
  &
  $B^* \Xi_c^*$ & $0$ & $-1$
  & $17^{+13}_{-11}$ & $7954^{+11}_{-13}$ & $\frac{1}{2}$ & B
  \\
  $B^* \Xi_c^*$ & $0$ & $-1$
  & $37^{+21}_{-18}$ & $7933^{+18}_{-21}$ & $\frac{3}{2}$ & A
  &
  $B^* \Xi_c^*$ & $0$ & $-1$
  & $29^{+18}_{-15}$ & $7942^{+15}_{-18}$ & $\frac{3}{2}$ & B
  \\
  $B^* \Xi_c^*$ & $0$ & $-1$
  & $17^{+13}_{-11}$ & $7954^{+11}_{-13}$ & $\frac{5}{2}$ & A
  &
  $B^* \Xi_c^*$ & $0$ & $-1$
  & $51^{+25}_{-23}$ & $7919^{+23}_{-25}$ & $\frac{5}{2}$ & B
  \\
  \hline
  $B \Xi_c^*-B_s \Sigma_c^*$ & $1$ & $-1$
  & $19^{+16}_{-13}$ & $7866^{+13}_{-16}$ & $\frac{3}{2}$ &A 
  &
  $B \Xi_c^*-B_s \Sigma_c^*$  & $1$ & $-1$
  & $27^{+19}_{-16}$ & $7858^{+17}_{-19}$ & $\frac{3}{2}$ & B
  \\
  $B^* \Xi_c^*-B_s^* \Sigma_c^*$ & $1$ & $-1$
  & $40^{+24}_{-21}$ & $7893^{+21}_{-24}$ & $\frac{1}{2}$ & A
  &
  $B^* \Xi_c^*-B_s^* \Sigma_c^*$ & $1$ & $-1$
  & $9.2^{+11.2}_{-8.1}$ & $7924.4^{+8.1}_{-11.2}$ & $\frac{1}{2}$ & B
  \\
  $B^* \Xi_c^*-B_s^* \Sigma_c^*$ & $1$ & $-1$ &
  $27^{+19}_{-16}$ & $7906^{+19}_{-16}$ & $\frac{3}{2}$ & A
  &
  $B^* \Xi_c^*-B_s^* \Sigma_c^*$ & $1$ & $-1$ &
  $20^{+16}_{-13}$ & $7914^{+13}_{-16}$ & $\frac{3}{2}$ & B
  \\
  $B^* \Xi_c^*-B_s^* \Sigma_c^*$ & $1$ & $-1$ &
  $9.2^{+11.2}_{-8.1}$ & $7924.4^{+8.1}_{-11.2}$ & $\frac{5}{2}$ & A
  &
  $B^* \Xi_c^*-B_s^* \Sigma_c^*$ & $1$ & $-1$ &
  $40^{+24}_{-21}$ & $7893^{+21}_{-24}$ & $\frac{5}{2}$ & B
  \\
  \hline
  $B \Omega_c^*-B_s \Xi_c^*$ & $\tfrac{1}{2}$ & $-2$
  & $12^{+16}_{-11}$ & $8000^{+11}_{-16}$ & $\frac{3}{2}$ & A
  &
  $B \Omega_c^*- B_s \Xi_c^*$ & $\tfrac{1}{2}$ & $-2$
  & $20^{+19}_{-16}$ & $7993^{+16}_{-19}$ & $\frac{3}{2}$ & B
  \\
  $B^* \Omega_c^*-B_s^* \Xi_c^*$ & $\tfrac{1}{2}$ & $-2$
  & $35^{+25}_{-22}$ & $8026^{+22}_{-25}$ & $\frac{1}{2}$ & A
  &
  $B^* \Omega_c^*-B_s^* \Xi_c^*$ & $\tfrac{1}{2}$ & $-2$
  & $3.9^{+10.7}_{-3.9}$ & $8057.1^{+3.9}_{-10.7}$ & $\frac{1}{2}$ & B
  \\
  $B^* \Omega_c^*-B_s^* \Xi_c^*$ & $\tfrac{1}{2}$ & $-2$
  & $22^{+20}_{-16}$ & $8040^{+16}_{-20}$ & $\frac{3}{2}$ & A
  &
  $B^* \Omega_c^*-B_s^* \Xi_c^*$ & $\tfrac{1}{2}$ & $-2$
  & $14^{+16}_{-12}$ & $8048^{+12}_{-16}$ & $\frac{3}{2}$ & B
  \\
  $B^* \Omega_c^*-B_s^* \Xi_c^*$ & $\tfrac{3}{2}$ & $-2$
  & $3.9^{+10.7}_{3.9}$ & $8057.1^{+3.9}_{-10.7}$ & $\frac{5}{2}$ & A
  &
  $B^* \Omega_c^*-B_s^* \Xi_c^*$ & $\tfrac{5}{2}$ & $-2$
  & $35^{+25}_{-22}$ & $8026^{+22}_{-25}$ & $\frac{5}{2}$ & B
  \\
  \hline \hline
  $B \Sigma_b^*$ & $\tfrac{1}{2}$ & \phantom{+}$0$
  & $46^{+21}_{-17}$ & $11065^{+17}_{-21}$ & $\frac{3}{2}$ & A
  &
  $B \Sigma_b^*$ & $\tfrac{1}{2}$ & \phantom{+}$0$
  & $57^{+23}_{-9}$ & $11055^{+9}_{-23}$ & $\frac{3}{2}$ & B
  \\
  $B^* \Sigma_b^*$ & $\tfrac{1}{2}$ & \phantom{+}$0$
  & $73^{+26}_{-22}$ & $11084^{+22}_{-26}$ & $\frac{1}{2}$ & A
  &
  $B^* \Sigma_b^*$ & $\tfrac{1}{2}$ & \phantom{+}$0$
  & $32^{+18}_{-14}$ & $11125^{+14}_{-18}$ & $\frac{1}{2}$ & B
  \\
  $B^* \Sigma_b^*$ & $\tfrac{1}{2}$ & \phantom{+}$0$
  & $57^{+23}_{-19}$ & $11100^{+19}_{-23}$ & $\frac{3}{2}$ & A
  &
  $B^* \Sigma_b^*$ & $\tfrac{1}{2}$ & \phantom{+}$0$
  & $47^{+21}_{-17}$ & $11110^{+17}_{-21}$ & $\frac{3}{2}$ & B
  \\
  $B^* \Sigma_b^*$ & $\tfrac{1}{2}$ & \phantom{+}$0$
  & $32^{+18}_{-14}$ & $11125^{+14}_{-18}$ & $\frac{5}{2}$ & A
  &
  $B^* \Sigma_b^*$ & $\tfrac{1}{2}$ & \phantom{+}$0$
  & $73^{+26}_{-22}$ & $11084^{+22}_{-26}$
  & $\frac{5}{2}$ & B
  \\
  \hline
  $B \Xi_b^*$ & $0$ & $-1$
  & $47^{+28}_{-24}$ & $11186^{+24}_{-28}$ & $\frac{3}{2}$ & A
  &
  $B \Xi_b^*$ & $0$ & $-1$
  & $54^{+30}_{-27}$ & $11176^{+27}_{-30}$ & $\frac{3}{2}$ & B
  \\
  $B^* \Xi_b^*$ & $0$ & $-1$
  & $73^{+35}_{-32}$ & $11205^{+32}_{-35}$ & $\frac{1}{2}$ & A
  &
  $B^* \Xi_b^*$ & $0$ & $-1$
  & $33^{+23}_{-19}$ & $11246^{+19}_{-23}$ & $\frac{1}{2}$ & B
  \\
  $B^* \Xi_b^*$ & $0$ & $-1$
  & $58^{+31}_{-27}$ & $11221^{+27}_{-31}$ & $\frac{3}{2}$ & A
  &
  $B^* \Xi_b^*$ & $0$ & $-1$
  & $47^{+28}_{-24}$ & $11231^{+24}_{-28}$ & $\frac{3}{2}$ & B
  \\
  $B^* \Xi_b^*$ & $0$ & $-1$
  & $33^{+23}_{-19}$ & $11246^{+19}_{-23}$ & $\frac{5}{2}$ & A
  &
  $B^* \Xi_b^*$ & $0$ & $-1$
  & $73^{+35}_{-32}$ & $11205^{+32}_{-35}$ & $\frac{5}{2}$ & B
  \\
  \hline
  $B \Xi_b^*-B_s \Sigma_b^*$ & $1$ & $-1$
  & $39^{+27}_{-22}$ & $11161^{+22}_{-27}$ & $\frac{3}{2}$ & A
  &
  $B \Xi_b^*-B_s \Sigma_b^*$  & $1$ & $-1$
  & $49^{+30}_{-26}$ & $11151^{+26}_{-30}$ & $\frac{3}{2}$ & B
  \\
  $B^* \Xi_b^*-B_s^* \Sigma_b^*$ & $1$ & $-1$
  & $65^{+35}_{-31}$ & $11183^{+31}_{-35}$ & $\frac{1}{2}$ & A
  &
  $B^* \Xi_b^*-B_s^* \Sigma_b^*$ & $1$ & $-1$ &
  $26^{+22}_{-17}$ & $11223^{+17}_{-22}$ & $\frac{1}{2}$ & B
  \\
  $B^* \Xi_b^*-B_s^* \Sigma_b^*$ & $1$ & $-1$ &
  $49^{+30}_{-26}$ & $11199^{+26}_{-30}$ & $\frac{3}{2}$ & A
  &
  $B^* \Xi_b^*-B_s^* \Sigma_b^*$ & $1$ & $-1$ &
  $40^{+27}_{-23}$ & $11208^{+23}_{-27}$ & $\frac{3}{2}$ & B
  \\
  $B^* \Xi_b^*-B_s^* \Sigma_b^*$ & $1$ & $-1$ &
  $26^{+22}_{-17}$ & $11223^{+17}_{-22}$ & $\frac{5}{2}$ & A
  &
  $B^* \Xi_b^*-B_s^* \Sigma_b^*$ & $1$ & $-1$ &
  $65^{+35}_{-31}$ & $11183^{+31}_{-35}$ & $\frac{5}{2}$ & B
  \\
  \hline
  $B \Omega_b^*-B_s \Xi_b^*$ & $\tfrac{1}{2}$ & $-2$ &
  $33^{+27}_{-22}$ & $11288^{+22}_{-27}$ & $\frac{3}{2}$ & A
  &
  $B \Omega_b^*- B_s \Xi_b^*$ & $\tfrac{1}{2}$ & $-2$
  & $43^{+30}_{-26}$ & $11278^{+26}_{-30}$ & $\frac{3}{2}$ & B
  \\
  $B^* \Omega_b^*-B_s^* \Xi_b^*$ & $\tfrac{1}{2}$ & $-2$
  & $60^{+35}_{-31}$ & $11309^{+31}_{-35}$ & $\frac{1}{2}$ & A
  &
  $B^* \Omega_b^*-B_s^* \Xi_b^*$ & $\tfrac{1}{2}$ & $-2$
  & $21^{+22}_{-17}$ & $11349^{+17}_{-22}$ & $\frac{1}{2}$ & B
  \\
  $B^* \Omega_b^*-B_s^* \Xi_b^*$ & $\tfrac{1}{2}$ & $-2$
  & $45^{+30}_{-26}$ & $11324^{+26}_{-30}$ & $\frac{3}{2}$ & A
  &
  $B^* \Omega_b^*-B_s^* \Xi_b^*$ & $\tfrac{1}{2}$ & $-2$
  & $35^{+27}_{-23}$ & $11334^{+23}_{-27}$ & $\frac{3}{2}$ & B
  \\
  $B^* \Omega_b^*-B_s^* \Xi_b^*$ & $\tfrac{3}{2}$ & $-2$
  & $21^{+22}_{-17}$ & $11349^{+17}_{-22}$ & $\frac{5}{2}$ & A
  &
  $B^* \Omega_b^*-B_s^* \Xi_b^*$ & $\tfrac{5}{2}$ & $-2$
  & $60^{+35}_{-31}$ & $11309^{+31}_{-35}$ & $\frac{5}{2}$ & B
  \\
  \hline \hline
\end{tabular}
\caption{
  Same as Table \ref{tab:HQS-partners-ccbar-bcbar} but for the $c \bar{b}$
  and $b \bar{b}$ sectors.
}
\label{tab:HQS-partners-cbbar-bbbar}
\end{table*}

\section{Compositeness of the pentaquarks}

Here we have described the pentaquarks as meson-baryon bound states,
which implicitly assumes that they are predominantly molecular or
composite in nature.
Yet, owing to the unspecified nature of the interaction binding the meson
and the baryon (which could have its origin in elementary components, e.g.
a five-quark compact core~\cite{Yamaguchi:2017zmn,Yamaguchi:2019seo}) and
the finite binding energy of these states, it is sensible to expect
that they will not be purely molecular.

From the EFT point of view, our assumption that the wave function of
a pentaquark only involves meson-baryon degrees of freedom is
expected to be valid up to $\mathcal{O}(Q/M)$ corrections:
\begin{eqnarray}
  | P_{QQ'} \rangle &=& | \mbox{meson-baryon} \rangle \times
  \left( 1 - \mathcal{O}(\frac{Q}{M}) \right) \nonumber \\
  &+& \mathcal{O}(\frac{Q}{M})\,| \mbox{compact} \rangle \, .
\end{eqnarray}
Here a caveat is in place: the wave function is not an observable and
as a consequence there will always remain a degree of ambiguity
on whether a particular state is composite or not
(or how composite it is).
In fact, the EFT framework usually does not rely on including new degrees of
freedom at subleading orders in the wave function to improve predictions.
Instead, it includes new contact-range operators acting on the degrees of
freedom already present, which means that compact components often
manifest as energy dependence.

Be it as it may, EFT can be used to derive
a dimensional estimation of the compositeness ($X_{\rm comp}$,
i.e. the probability of the meson-baryon component) of the pentaquarks
\begin{eqnarray}
  X_{\rm comp}^{\rm dim}(P_{QQ'}) &=& 1 - \mathcal{O}(\frac{Q}{M})
  = \frac{1}{1 + \mathcal{O}(\frac{Q}{M})}
  \nonumber \\
  &\approx& \frac{1}{1 + x_c\,\frac{\sqrt{2 \mu B_2}}{m_{\rho}}}
  + \mathcal{O}(\frac{Q}{M})
  \, , \label{eq:X_comp_dim}
\end{eqnarray}
where we have reordered the terms in order to obtain an expression
that is suitable when $Q/M$ is not small (i.e. when the binding
energy is closer to the limit at which the EFT will fail,
so we only have $Q/M < 1$ but not $Q/M \ll 1$).
In the second line we have particularized for the choice
$Q = \gamma_2 = \sqrt{2\mu B_2}$ and $M = m_{\rho}$,
where $x_c$ is a numerical constant of $\mathcal{O}(1)$
for which we will choose $x_c = 1$.
This yields a compositeness of around $X_{\rm comp}^{\rm dim} = (0.85,0.78,0.88)$
for the $P_{c1}$, $P_{c2}$ and $P_{c3}$ pentaquarks in the $c\bar{c}$ sector,
$(0.76, 0.70,0.79)$ and $(0.71,0.67,0.75)$ for $\bar{c} b$ and $c\bar{b}$,
respectively, while merely a value of $(0.60, 0.56,0.63)$
for their $b \bar{b}$ counterparts.
As a comparison, for the deuteron ($\gamma_2 = 45\,{\rm MeV}$)
we will obtain a compositeness of $0.94$, compatible
with a pure molecular interpretation.
Yet, we remind that these estimates are purely based on a comparison of
scales and are not very precise.
This is illustrated by the numerical factor $x_c$ in Eq.~(\ref{eq:X_comp_dim}),
where by taking $x_c = 1/2$ or $x_c = 2$ instead of $x_c = 1$ (all of which
are $\mathcal{O}(1)$), the compositeness will change
by a factor of order $Q/M$.

Actually, there is a rich literature dealing with ways of quantifying
the compositeness of a state~\cite{Weinberg:1962hj,Weinberg:1963zza,Weinberg:1965zz,Baru:2010ww,Hyodo:2013iga,Sekihara:2014kya,Kamiya:2015aea,Kamiya:2016oao,Sekihara:2016xnq,Matuschek:2020gqe,Song:2022yvz,Albaladejo:2022sux},
which we can use to obtain a refined estimation of $X_{\rm comp}$.
They began with the compositeness criterion proposed by Weinberg~\cite{Weinberg:1962hj,Weinberg:1963zza,Weinberg:1965zz},
which can be written as
\begin{eqnarray}
  {X}^{\rm W}_{\rm comp} = \sqrt{\frac{1}{1 - 2 \frac{r_0}{a_0}}} \, ,
\end{eqnarray}
where $a_0$ and $r_0$ are the scattering length~\footnote{In our convention,
  for attractive potentials $a_0 < 0$ in the absence of bound states and
  $a_0 > 0$ when there is one bound state.} and effective range and
which showed in a model-independent way that the deuteron is
probably composite.
It actually returns $X_{\rm comp} > 1$ for the deuteron, which indicates we are
using the previous formula beyond its domain of validity ($r_0 < 0$ for
obtaining $X_{\rm comp} < 1$ for a bound state, not to mention
that there will be corrections coming from the range of
the interaction, as already pointed in~\cite{Weinberg:1965zz}),
but this result is usually interpreted as molecular.
The bottom-line though is that the Weinberg criterion relies heavily
on the sign of the effective range of the purported components of the state:
if positive (negative) the state will be predominantly composite (elementary).
As a consequence the application of this criterion will lead to the conclusion
that the pentaquarks we are dealing with here are mostly molecular.
This however will be an artifact of the formalism we are using:
our LO calculation automatically generates a positive effective
range, which is a consequence of the dynamics we are using~\footnote{
  Only at $\rm NLO$ will we be able to obtain a negative effective range,
  as it is at this order that energy and momentum dependent corrections
  to the contact-range potential enter.
  Unfortunately this calculation implies new couplings, the calibration of which
  require meson-baryon scattering data that are not available at the moment.}.
Besides, even though it is evident that the energy dependence of a compact core
coupled to a two-hadron system is such that it will generate a negative
effective range, a sufficiently short-ranged potential combined with
a large binding energy implies a sizable superposition of the hadrons
and, owing to their finite size, also a degree of non-compositeness.
From this and other arguments, extensions of the Weinberg criterion have
been proposed that apply to situations different from a bound state with
negative effective range~\cite{Baru:2010ww,Hyodo:2013iga,Sekihara:2014kya,Kamiya:2015aea,Kamiya:2016oao,Sekihara:2016xnq,Matuschek:2020gqe,Song:2022yvz,Albaladejo:2022sux}.

A recent proposal of a model-independent estimation of the compositeness
of a state is the following~\cite{Matuschek:2020gqe}
\begin{eqnarray}
  \tilde{X}_{\rm comp} = \sqrt{\frac{1}{1 + 2 | \frac{r_0}{a_0} |}} \, ,
  \label{eq:X_tilde_comp}
\end{eqnarray}
which returns $\tilde{X}_{\rm comp} < 1$,
where the calculation of $a_0$ and $r_0$ for our contact-range theory
is explained in Appendix \ref{app:ERE}.
This criterion would provide a compositeness of
$\tilde{X}_{\rm comp} = (0.73,0.67,0.76)$ for each of the three LHCb pentaquarks
(i.e. $P_{c1}$, $P_{c2}$, $P_{c3}$), $(0.66,0.62,0.68)$ and $(0.63,0.61,0.65)$
for the $\bar{c} b$ and $c\bar{b}$ ones and $(0.59,0.57,0.60)$
for the hidden-bottom $P_{b1}$, $P_{b2}$ and $P_{b3}$ pentaquarks.
However, the problem here is that we are using a LO EFT description with
only one parameter (the binding energy), which means that the value of
the effective range thus obtained is only a dimensional estimation
within our EFT.
For comparison the compositeness of the deuteron ($a_0 = 5.419\,{\rm fm}$,
$r_0 = 1.753\,{\rm fm}$~\cite{deSwart:1995ui}) with this criterion
will be $0.78$, but in this case there is plenty of experimental
information available about neutron-proton scattering, i.e.
$a_0$ and $r_0$ are well-known.

Regardless of the specific criterion used to estimate compositeness
(after all, the wave function is not an observable),
it seems that in general the hidden-charm pentaquarks are less composite
than the deuteron, and as we move into heavier flavor sectors
their compositeness reduces further.
This is in turn compatible with the observation that the EFT description
is less convergent and has larger uncertainties for two-body systems
with larger binding energies.
Thus, as binding increases with the reduced mass,
we expect compositeness to decrease accordingly.

\section{Flavor symmetry and non-molecular explanations}

The present predictions have been done under the assumption
that the hidden-charm pentaquarks are molecular.
But, as a matter of fact, the light- and heavy-flavor symmetries
we have used here are expected to apply to other light-heavy hadrons as well,
independently of their nature (though the uncertainties stemming
from the violations of these symmetries could be very different).
For instance, the existence of this type of pentaquark multiplets
has been predicted \mpv{in the compact}~\cite{Santopinto:2016pkp} and
hadroquarkonium pictures~\cite{Eides:2015dtr,Ferretti:2020ewe}.
\mpv{Theoretical explorations in the previous pictures have been mostly
  concentrated in the hidden-charm sector, where the mass splittings of
  the octet [$m(P_c^\Lambda) - m(P_c^N)$, $m(P_c^\Sigma) - m(P_c^N)$ and
    $m(P_c^\Xi) - m(P_c^N)$] are $141$, $205$ and $315\,{\rm MeV}$
  for compacts pentaquarks~\cite{Santopinto:2016pkp}
  and $150$, $217$ and $327\,{\rm MeV}$
  for hadrocharmonia~\cite{Eides:2017xnt}.
  These mass splittings happen to be larger than for molecular pentaquarks
  ($125$, $105$ and $232\,{\rm MeV}$) and might provide a way
  to distinguish their nature if they are observed.
  For the hidden-bottom sector there are indeed predictions of $P_b^N$
  pentaquarks in the local hidden-gauge approach of Ref.~\cite{Xiao:2013jla}
  and in models considering a five-quark core and
  pion exchanges~\cite{Yamaguchi:2017zmn}.
}
It is plausible that other theoretical models of $Q'\bar{Q}$ pentaquarks will
lead to analogous predictions for their flavor partners,
as these predictions are constrained by symmetry principles
(instead of the details of the dynamics, which will matter
for how the spectrum is organized in terms of
quantum numbers, spin-spin splitting, etc.).
Recent calculations of $qq s Q' \bar{Q}$ pentaquarks
in the hadroquarkonium~\cite{Ferretti:2020ewe} and
chiral quark models~\cite{Zhang:2020cdi} provide
further support for this conjecture.

\section{Summary}

The observation of the LHCb hidden-charm pentaquarks
in combination with SU(3)- and heavy-flavor symmetries leads
to the prediction of a series of flavor partners.
In particular, \mpv{pentaquarks (molecular and
  non-molecular~\cite{Santopinto:2016pkp} alike)}
are expected to form a light-flavor
octet reminiscent of the light{-}baryon octet and are also expected to appear
in the $c {\bar b}$, ${\bar c} {b}$ and $b {\bar b}$ sectors
as well as in the original hidden-charm sector
where they have been discovered.
We denote these pentaquarks as $P_{Q' \bar{Q}}^N$, $P_{Q' \bar{Q}}^{\Lambda}$,
$P_{Q' \bar{Q}}^{\Sigma}$, $P_{Q' \bar{Q}}^{\Xi}$, with the superscript and
subscript referring to their light- and heavy-quark structure,
respectively (which we shorten to $P_Q^N$, $P_Q^{\Lambda}$, $P_Q^{\Sigma}$
and $P_Q^{\Xi}$ when the heavy flavors coincide $Q' = Q$,
i.e. for hidden-flavor).
For predicting their masses, we have made use of a contact-range theory
with a natural cutoff in the range $\Lambda = (0.5-1.0)\,{\rm GeV}$.
Among the predictions, it is worth noticing the existence of
five-flavor pentaquarks, i.e. pentaquarks containing all
the five flavors that hadronize ($P_{c \bar{b}}^{\Lambda}$, $P_{{\bar c} {b}}^{\Lambda}$,
$P_{c \bar{b}}^{\Sigma}$, $P_{{\bar c} {b}}^{\Sigma}$ in our notation)
in the $7770-7910\,{\rm MeV}$ region.
The five-flavor pentaquarks could be detected via
their $B_c^{\pm} \Lambda$ and $B_c^{\pm} \Sigma$ decays.

The predictions made in this work assume the LHCb pentaquarks
to be meson-baryon bound states \mpv{the dynamics of which
  can be described in terms of a contact-range theory.
  It is worth noticing that the applicability of this description
  decreases with increasing binding energy, as this implies
  pentaquarks that are less composite, and with heavier
  reduced masses owing to the model-dependent
  nature of HFS~\cite{Baru:2018qkb}.
  This is reflected in the larger uncertainties,
  particularly in the hidden-bottom sector.} Yet,
  it is sensible to expect these predictions to be more dependent
  on the general symmetry principles we have applied than
  on the details of the dynamics generating
  the pentaquarks, \mpv{e.g. models with a compact five-quark core coupled
  to the meson-baryon degrees of freedom do reproduce the hidden-charm
  pentaquarks~\cite{Yamaguchi:2019seo} and also predict
  the hidden-bottom ones~\cite{Yamaguchi:2017zmn},
  giving credence to the aforementioned conjecture.
}
Thus it might be the case that the light- and heavy-flavor symmetry
partners of the hidden-charm pentaquarks exist irrespective of
the binding mechanism, though the details of the spectrum
will be different than in the molecular case.

\section*{Acknowledgments}
This work is partly supported by the National Natural Science Foundation of
China under Grants No.11735003 and No. 11975041, the fundamental Research
Funds for the Central Universities and the Thousand
Talents Plan for Young Professionals.
M.P.V. thanks the IJCLab of Orsay, where part of this work was done,
for its hospitality.

\appendix
\section{Heavy-quark mass dependence of the binding energy}
\label{app:binding-mass}

Here we consider the variation of the binding energy of a heavy hadron
molecule with respect to the heavy-quark mass.
If the potential between two heavy hadrons does not depend
on the heavy-quark mass, it can be shown that
the binding energy increases with the heavy-quark mass
(in agreement with naive expectations).

At leading order in the $1/m_Q$ expansion,
we can write the Schr\"odinger equation for a heavy hadron molecule as follows
\begin{eqnarray}
- \nabla^2 \Psi_{Q}(\vec{r}) + 2\,\mu_Q\,V_Q(\vec{r})\,
\Psi_{Q}(\vec{r}) &=& -2\,\mu_Q B_Q \Psi_{Q}(\vec{r}) \, , \nonumber \\
\end{eqnarray}
where the subindex ${}_Q$ indicates the dependence (explicit and implicit)
on the heavy-quark mass, $\Psi_Q$ is the wave function,
$\mu_Q$ the reduced mass of the molecule, $V_Q$ the potential
and $B_Q$ the two-body binding energy.
We can construct a Wronskian identity for the Schr\"odinger equation
at two different heavy-quark masses as follows
\begin{eqnarray}
- \left( \Psi_{Q'} \nabla^2 \Psi_{Q} - \Psi_{Q} \nabla^2 \Psi_{Q'}\right)
\quad && \nonumber \\
+ 2\,\left(\mu_Q\,V_Q - \mu_{Q'} V_{Q'} \right)\,
\Psi_{Q}\,\Psi_{Q'}
&& \nonumber \\
= - 2\,(\mu_Q B_Q - \mu_{Q'} B_{Q'})\,
\Psi_{Q}\,\Psi_{Q'} \, , &&
\end{eqnarray}
where, again, ${}_Q$ and ${}_{Q'}$ represent the different quantities we are
considering at $m_Q$ and $m_{Q'}$, respectively.
The Wronskian identity can be integrated, leading to
\begin{eqnarray}
&& 2\,\int d^3\vec{r}\,\left(\mu_Q\,V_Q - \mu_{Q'} V_{Q'} \right)\,
\Psi_{Q}(\vec{r})\,\Psi_{Q'}(\vec{r}) = \nonumber \\
&& \qquad - 2\,(\mu_Q B_Q - \mu_{Q'} B_{Q'})\,
\int d^3\vec{r}\,\Psi_{Q}(\vec{r})\,\Psi_{Q'}(\vec{r})
\end{eqnarray}
where the kinetic term disappears because it is exactly differentiable and
can be rewritten as a surface term, which vanishes if we consider bound
state solutions.
Now we will consider a small change in the heavy-quark mass,
which we can symbolically indicate by
\begin{eqnarray}
Q' = Q + \delta\,Q \, .
\end{eqnarray}
We can deduce that
\begin{eqnarray}
\int d^3\vec{r} \, \Psi_{Q}(\vec{r})\,\Psi_{Q'}(\vec{r}) &=& 1 + (\delta\,Q)^2
\end{eqnarray}
which is a consequence of the normalization of the wave function
(i.e. $\langle \Psi_Q | \Psi_Q \rangle = \langle \Psi_{Q'} | \Psi_{Q'} \rangle = 1$, which is why the $\delta Q$ term vanishes).
If we assume that the potential does not depend on the heavy-quark mass, i.e.
$V_Q = V_{Q'}$, we can use the previous result to prove that
\begin{eqnarray}
2\,\delta \mu_Q\,\langle V_Q \rangle = - 2\,\delta(\mu_Q B_Q) \, ,
\end{eqnarray}
which we can differentiate to obtain
\begin{eqnarray}
\langle V_Q \rangle = - B_Q - \mu_Q\,\frac{\partial B_Q}{\partial \mu_Q} \, .
\end{eqnarray}
If we take into account
\begin{eqnarray}
  \langle T_Q \rangle + \langle V_Q \rangle = - B_Q \, ,
\end{eqnarray}
where $\langle T_Q \rangle \geq 0$ is the kinetic energy of the heavy molecule,
we can rewrite the binding energy dependence on the reduced mass as
\begin{eqnarray}
\langle T_Q \rangle = \mu_Q\,\frac{\partial B_Q}{\partial \mu_Q}
\end{eqnarray}
or, equivalently
\begin{eqnarray}
\frac{\partial B_Q}{\partial \mu_Q} \geq 0 \, ,
\end{eqnarray}
as a consequence of the fact that the kinetic energy is positive.
That is, the system will become more bound the heavier the mesons
(this is a model-independent result).
What is difficult (and model-dependent) is to determine by what amount.
Finally, we notice that including a heavy-quark mass dependence of the type
$V_Q = V_0 + \frac{1}{m_Q}\,V_1 + \dots$ in the potential does only
induce $1/m_Q$ corrections to the previous relation,
which can be safely neglected in the heavy-quark mass limit.

\section{Calculation of the effective range expansion parameters}
\label{app:ERE}

The evaluation of the different compositeness conditions available
in the literature usually require the effective range parameters as input.
Here we briefly explain how to calculate them.
We begin by writing down the relation between the effective range
expansion and the on-shell T-matrix ($T_{\rm os}$):
\begin{eqnarray}
  -\frac{2\pi}{\mu}\,{\rm Re}\,\left[ \frac{1}{T_{\rm os}(k)} \right] =
  - \frac{1}{a_0} + \frac{1}{2}\,r_0\,k^2 + \sum_{n=2}^{\infty} v_n k^{2n} \, ,
\end{eqnarray}
where $a_0$ is the scattering length, $r_0$ the effective range, $v_n$
the shape parameters, $k$ the center-of-mass momentum and $\mu$
refers to the reduced mass of the two-body system.
For attractive potentials, the previous convention implies $a_0 < 0$ if
there is no bound state (or an even number of bound states) and
$a_0 > 0$ if there is an odd number of bound states.
The on-shell T-matrix corresponds to the following matrix element of
the full T-matrix
\begin{eqnarray}
  T_{\rm os}(k) = \langle k | T(k) | k \rangle \, ,
\end{eqnarray}
where $T$ obeys the Lippmann-Schwinger equation, which for scattering states
takes the form
\begin{eqnarray}
  T = V + V G_0(E + i \epsilon) T \, ,
\end{eqnarray}
with $G_0(E) = 1 / (E - H_0)$ the resolvent operator and
$E = k^2 / 2 \mu$ the center-of-mass energy of
the system.
If we consider a regularized contact-range of the type
\begin{eqnarray}
  \langle p' | V_C | p \rangle =
  C(\Lambda)\,f(\frac{p'}{\Lambda})\,f(\frac{p}{\Lambda})
  \, ,
\end{eqnarray}
then the explicit solution of the Lippmann-Schwinger equation
for the on-shell T-matrix reads
\begin{eqnarray}
  {\rm Re}\left[ \frac{1}{T_{\rm os}(k)} \right] = \frac{1}{C(\Lambda)} -
  \frac{\mu}{\pi^2}\,\mathcal{P}\,\int_0^{\infty}\,
  \frac{p^2 dp}{k^2 + i \epsilon - p^2}\,f^2(\frac{p}{\Lambda}) \, ,
  \nonumber \\
\end{eqnarray}
where $\mathcal{P}$ denotes the principal value of the integral.
By expanding in powers of the center-of-mass momentum,
we arrive at
\begin{eqnarray}
  \frac{1}{a_0} &=& \frac{2\pi}{\mu}\,\frac{1}{C(\Lambda)} +
  \frac{2}{\pi}\,\int_0^{\infty} dp \, f^2(\frac{p}{\Lambda}) \, , \\
  r_0 &=& - \frac{4}{\pi}\,\int_0^{\infty} \,\frac{dp}{p^2}\,
  \left( f^2(\frac{p}{\Lambda}) - f^2(0) \right) \, ,
\end{eqnarray}
where we can appreciate that at ${\rm LO}$ in our contact-range theory
$r_0$ depends solely on the regulator and cutoff, i.e. EFT merely
provides a dimensional estimation of its size.
If we particularize for our choices of regulator function, cutoff and couplings,
we will obtain the values of $a_0$ and $r_0$ that we have used
as input for Eq.~(\ref{eq:X_tilde_comp}).

%

\end{document}